\PassOptionsToPackage{colorlinks=true,hypertexnames=false}{hyperref}
\PassOptionsToPackage{usenames,dvipsnames}{xcolor}

  \documentclass[acmsmall,screen,nonacm,natbib=false]{acmart}

\definecolor{orcidgreen}{HTML}{A6CE39}
\usepackage{fontawesome5}

\newcommand{\orcidID}{0000-0002-2748-9091}

\makeatletter
\def\@typeset@author@line{%
  \andify\@currentauthors\par\noindent
  \@currentauthors\def\@currentauthors{}%
  \,\href{https://orcid.org/\orcidID}{\raisebox{.2em}{\normalsize\color{orcidgreen}\faOrcid}}%
  \ifx\@currentaffiliations\@empty\else
    \andify\@currentaffiliations
      \unskip, {\@currentaffiliations}\par
  \fi
  \def\@currentaffiliations{}}
\makeatother

\makeatletter
\gdef\@acmBooktitle{}
\makeatother

\usepackage{ifthen}

\makeatletter
\fancypagestyle{firstpage}{%
  \lhead{\footnotesize Prepared for submission to \textit{\journalname}}
  \ifthenelse{\equal{\@date}{}}{
    \rhead{\footnotesize \texttt{\reportnumber}}
  }{
    \rhead{\footnotesize \texttt{\reportnumber}\\\@date}
  }
}

\makeatother
\let\oldmaketitle\maketitle
\renewcommand{\maketitle}{\oldmaketitle \thispagestyle{firstpage}}

\AtBeginDocument{\hypersetup{linkcolor=Blue,urlcolor=Maroon,citecolor=Fuchsia}}
\AtBeginDocument{}

\usepackage[style=ACM-Reference-Format,citestyle=numeric-comp,backend=bibtex,sorting=none]{biblatex}

\usepackage[utf8]{inputenc}
\usepackage{amsthm}
\usepackage[T1]{fontenc}
\usepackage{mathtools}
\usepackage{bm}
\usepackage{bbm}
\usepackage[normalem]{ulem}
\usepackage{caption}
\usepackage{placeins}
\usepackage{braket}
\usepackage{algorithm,algorithmic}
\usepackage{tikz}
\usepackage{quantikz}
\usetikzlibrary{positioning}
\usetikzlibrary{shapes}
\usetikzlibrary{calc}

\allowdisplaybreaks

\usepackage{refcount}
\newcommand{\footref}[1]{\hyperref[#1]{\footnotemark[\getrefnumber{#1}]}}

\usepackage{tabularx}
\newcolumntype{P}[1]{>{\centering\arraybackslash}p{#1}} 
\newcolumntype{M}[1]{>{\centering\arraybackslash}m{#1}} 
\newcolumntype{B}[1]{>{\centering\arraybackslash}b{#1}} 

\usepackage{dcolumn}
\newcolumntype{.}{D{.}{.}{-1}} 

\newcommand{\fref}[1]{\hyperref[#1]{Figure~\ref*{#1}}}
\newcommand{\Fref}[1]{\hyperref[#1]{Figure~\ref*{#1}}}
\newcommand{\sref}[1]{\hyperref[#1]{Section~\ref*{#1}}}
\newcommand{\Sref}[1]{\hyperref[#1]{Section~\ref*{#1}}}
\newcommand{\tref}[1]{\hyperref[#1]{Table~\ref*{#1}}}
\newcommand{\Tref}[1]{\hyperref[#1]{Table~\ref*{#1}}}
\newcommand{\aref}[1]{\hyperref[#1]{Appendix~\ref*{#1}}}
\newcommand{\Aref}[1]{\hyperref[#1]{Appendix~\ref*{#1}}}
\newcommand{\thref}[1]{\hyperref[#1]{Theorem~\ref*{#1}}}
\newcommand{\Thref}[1]{\hyperref[#1]{Theorem~\ref*{#1}}}
\newcommand{\alref}[1]{\hyperref[#1]{Algorithm~\ref*{#1}}}
\newcommand{\Alref}[1]{\hyperref[#1]{Algorithm~\ref*{#1}}}

\let\oldPsi\Psi
\renewcommand{\Psi}{\bm{\oldPsi}}
\newcommand{\Udif}{\mathbf{S}_{\Psi_0}}
\newcommand{\Uphi}{\mathbf{U}_\varphi}
\newcommand{\Uiter}{\mathbf{Q}_\mathrm{iter}}
\newcommand{\I}{\mathbf{I}}
\renewcommand{\U}{\mathbf{U}}
\renewcommand{\S}{\mathbf{S}}
\newcommand{\Alpha}{\bm{\alpha}}
\newcommand{\Beta}{\bm{\beta}}
\newcommand{\Eta}{\bm{\eta}}

\newcommand{\bigket}[1]{\big|#1\big\rangle}

\newcommand{\Ueo}{Q_\mathrm{evenodd}}

\newcommand{\RE}{\mathrm{Re}}
\newcommand{\IM}{\mathrm{Im}}

\newcommand{\qwmult}{\qwbundle[alternate]{}}

\makeatletter
\newcommand{\vast}{\bBigg@{3}}
\newcommand{\Vast}{\bBigg@{4}}
\makeatother

\newcommand{\Grover}{10.1145/237814.237866}

\begin{document}

\title[Non-Boolean Quantum Amplitude Amplification and Quantum Mean Estimation]{Non-Boolean Quantum Amplitude Amplification and Quantum Mean Estimation}

\newcommand{\journalname}{ACM Transactions on Quantum Computing}
\newcommand{\reportnumber}{FERMILAB-PUB-21-018-QIS}
\date{February 09, 2021}

\author{Prasanth Shyamsundar}
\email{prasanth@fnal.gov}
\orcid{0000-0002-2748-9091}
\affiliation{%
  \institution{Fermi National Accelerator Laboratory}
  \department{Fermilab Quantum Institute}
  \streetaddress{PO Box 500}
  \city{Batavia}
  \state{Illinois}
  \postcode{60510-0500}
  \country{USA}
}

\begin{abstract}
This paper generalizes the quantum amplitude amplification and amplitude estimation algorithms to work with non-boolean oracles. The action of a non-boolean oracle $U_\varphi$ on an eigenstate $\Ket{x}$ is to apply a state-dependent phase-shift $\varphi(x)$. Unlike boolean oracles, the eigenvalues $\exp(i\varphi(x))$ of a non-boolean oracle are not restricted to be $\pm 1$. Two new oracular algorithms based on such non-boolean oracles are introduced. The first is the non-boolean amplitude amplification algorithm, which preferentially amplifies the amplitudes of the eigenstates based on the value of $\varphi(x)$. Starting from a given initial superposition state $\Ket{\psi_0}$, the basis states with lower values of $\cos(\varphi)$ are amplified at the expense of the basis states with higher values of $\cos(\varphi)$. The second algorithm is the quantum mean estimation algorithm, which uses quantum phase estimation to estimate the expectation $\Braket{\psi_0|U_\varphi|\psi_0}$, i.e., the expected value of $\exp(i\varphi(x))$ for a random $x$ sampled by making a measurement on $\Ket{\psi_0}$. It is shown that the quantum mean estimation algorithm offers a quadratic speedup over the corresponding classical algorithm. Both algorithms are demonstrated using simulations for a toy example.
Potential applications of the algorithms are briefly discussed.

\end{abstract}



\maketitle

\section{Introduction} \label{sec:intro}

Grover's algorithm, introduced in Ref.~\cite{\Grover}, is a quantum search algorithm for finding the unique input $x_\mathrm{good}$ that satisfies
\begin{equation}
 f_\mathrm{bool}(x_\mathrm{good}) = 1\,, \label{eq:winner} 
\end{equation}
for a given boolean function $f_\mathrm{bool}: \{0, 1, \dots, N-1\} \rightarrow \{0, 1\}$. Such an input $x_\mathrm{good}$ satisfying \eqref{eq:winner} is referred to as the \emph{winning} input of $f_\mathrm{bool}$. Grover's original algorithm has also been adapted to work with boolean functions with multiple winning inputs \cite{multiplewinning}, where the goal is to find any one of the winning inputs.

An important generalization of Grover's algorithm is the amplitude amplification algorithm \cite{595153,Grover:1997ch,QAAAE}, in which the function $f_\mathrm{bool}$ is accessed through a boolean quantum oracle $\hat{U}_{\!f_\mathrm{bool}}$ that acts on the orthonormal basis states $\Ket{0},\dots,\Ket{N-1}$ as follows:
\begin{equation} \label{eq:bool_oracle}
 \hat{U}_{\!f_\mathrm{bool}} \Ket{x} = \begin{cases}
  -\Ket{x}, \qquad & \text{if } f_\mathrm{bool}(x) = 1\,,\\
  +\Ket{x}, \qquad & \text{if } f_\mathrm{bool}(x) = 0\,.
 \end{cases}
\end{equation}
In this way, the oracle marks the winning states by flipping their phase (shifting the phase by $\pi$). Given a superposition state $\Ket{\psi_0}$, the goal of the amplitude amplification algorithm is to amplify the amplitudes (in the superposition state) of the winning states. The algorithm accomplishes this iteratively, by initializing a quantum system in the state $\Ket{\psi_0}$ and performing the operation $S_{\psi_0}\,\hat{U}_{\!f_\mathrm{bool}}$ on the system during each iteration, where
\begin{equation}
 S_{\psi_0} \equiv 2\Ket{\psi_0}\Bra{\psi_0} - I\,. \label{eq:S_onereg}
\end{equation}
Here, $I$ is the identity operator. Performing a measurement on the system after the iterative amplification process results in one of winning states with high probability. 
Grover's original algorithm is a special case of the amplitude amplification algorithm, where a) the uniform superposition state $\Ket{s}$, given by
\begin{equation} \label{eq:udif_onereg}
 \Ket{s} = \frac{1}{\sqrt{N}} \sum_{x=0}^{N-1}\, \Ket{x}\,,
\end{equation}
is used as the initial state $\Ket{\psi_0}$ of the system, and b) there is exactly one winning input.

Closely related to the amplitude amplification algorithm is the amplitude estimation algorithm \cite{10.1007/BFb0055105,QAAAE}. It combines ideas from the amplitude amplification algorithm and the quantum phase estimation (QPE) algorithm \cite{doi:10.1098/rspa.1998.0164} to estimate the probability that making a measurement on the initial state $\Ket{\psi_0}$ will yield a winning input. If the uniform superposition state $\Ket{s}$ is used as $\Ket{\psi_0}$, the amplitude estimation algorithm can help estimate the number of winning inputs of $f_\textrm{bool}$---this special case is also referred to as the quantum counting algorithm \cite{10.1007/BFb0055105}.

The amplitude amplification algorithm and the amplitude estimation algorithm have a wide range of applications, and are important primitives that feature as subroutines in a number of other quantum algorithms \cite{10.1145/261342.261346,10.1137/040605072,10.1007/BFb0054319,10.1007/s11128-017-1600-4,10.1007/978-3-540-78773-0_67,7016940,Yanhu,HOGG2000181,10.1145/301250.301349,Gong,zeng2}. 
The amplitude amplification algorithm can be used to find a winning input to $f_\mathrm{bool}$ with $\mathcal{O}\big(\sqrt{N}\,\big)$ queries\footnote{The $\mathcal{O}(\sqrt{N})$ and $\mathcal{O}(N)$ scalings for the quantum and classical algorithms, respectively, hold assuming that the number of winning states does not scale with $N$.\label{foot:noscale}} of the quantum oracle, regardless of whether the number of winning states is a priori known or unknown \cite{multiplewinning}. This represents a quadratic speedup over classical algorithms, which require $\mathcal{O}(N)$ evaluations\footref{foot:noscale} of the function $f_\mathrm{bool}$. Similarly, the amplitude estimation algorithm also offers a quadratic speedup over the corresponding classical approaches \cite{QAAAE}. The quadratic speedup due to the amplitude amplification algorithm has been shown to be optimal for oracular quantum search algorithms \cite{doi:10.1137/S0097539796300933}.

A limitation of the amplitude amplification and estimation algorithms is that they work only with boolean oracles, which classify the basis states as good and bad. On the other hand, one might be interested in using these algorithms in the context of a non-boolean function of the input $x$. In such situations, the typical approach is to create a boolean oracle from the non-boolean function, by using a threshold value of the function as the decision boundary---the winning states are the ones for which the value of the function is, say, less than the chosen threshold value \cite{nonbool1,nonbool2}. In this way, the problem at hand can be adapted to work with the standard amplitude amplification and estimation algorithms. Another approach is to adapt the algorithms to work directly with non-boolean functions \cite{HOGG2000181}.

This paper generalizes the amplitude amplification and estimation algorithms to work with quantum oracles for non-boolean functions. In particular, a) the \textbf{boolean} amplitude amplification algorithm of Ref.~\cite{QAAAE} is generalized to the \textbf{non-boolean} amplitude amplification algorithm, and b) the amplitude estimation algorithm of Ref.~\cite{QAAAE} is generalized to the quantum mean estimation algorithm. Henceforth, the qualifiers ``boolean'' and ``non-boolean'' will be used whenever necessary, in order to distinguish between the different versions of the amplitude amplification algorithm. The rest of this section introduces non-boolean oracles and describes the goals of the two main algorithms of this paper.




\subsection{Oracle for a Non-Boolean Function}
The behavior of the boolean oracle $\hat{U}_{\!f_\mathrm{bool}}$ in \eqref{eq:bool_oracle} can be generalized to non-boolean functions by allowing the oracle to perform arbitrary phase-shifts on the different basis states. More concretely, let $\varphi: \{0,1,\dots,N-1\} \rightarrow \mathbb{R}$ be a real-valued function, and let $U_\varphi$ be a quantum oracle given by
\begin{equation}
 U_\varphi \equiv \sum_{x=0}^{N-1}e^{i\varphi(x)}\Ket{x}\Bra{x}\,.\label{eq:gen_oracle_def}
\end{equation}
The actions of the oracle $U_\varphi$ and its inverse $U_\varphi^\dagger$ on the basis states $\Ket{0},\dots,\Ket{N-1}$ are given by
\begin{alignat}{3}
 U_\varphi \Ket{x} &= e^{+i\varphi(x)} &&\Ket{x} &&= 
 \Big[\cos\!\big(\varphi(x)\big) + i \sin\!\big(\varphi(x)\big)\Big]\Ket{x}\,, \label{eq:gen_oracle}\\
 U^\dagger_\varphi \Ket{x} &= e^{-i\varphi(x)} &&\Ket{x} &&= \Big[\cos\!\big(\varphi(x)\big) - i \sin\!\big(\varphi(x)\big)\Big]\Ket{x}\,. \label{eq:inv_gen_oracle}
\end{alignat}

\subsection{Goal of the Non-Boolean Amplitude Amplification Algorithm}
Given an oracle $U_\varphi$ and an initial state $\Ket{\psi_0}$, the goal of the non-boolean amplitude amplification algorithm introduced in this paper is to preferentially amplify the amplitudes of the basis states $\Ket{x}$ with lower values of $\cos\!\big(\varphi(x)\big)$, at the expense of the amplitude of states with higher values of $\cos\!\big(\varphi(x)\big)$. Depending on the context in which the algorithm is to be used, a different function of interest $f$ (which is intended to guide the amplification) can be appropriately mapped onto the function $\varphi$. For example, if the range of $f$ is $[0,1]$ and one intends to amplify the states with higher values of $f$, then \eqref{eq:map_options} shows two different options for formulating the problem in terms of $\varphi$.
\begin{equation}
 \varphi(x) = \pi f(x)\qquad \text{or}\qquad \varphi(x) = \arccos\!\big(1-2f(x)\big)\,. \label{eq:map_options}
\end{equation}
In both these cases, $\cos(\varphi)$ is monotonically decreasing in $f$.

The connection between the boolean and non-boolean amplitude amplification algorithms can be seen as follows: If either of the two options in \eqref{eq:map_options} is used to map a boolean function $f_\mathrm{bool}$ onto $\varphi_\mathrm{bool}$, then
\begin{equation}
 \varphi_\mathrm{bool}(x) = \begin{cases}
  \pi\,,\qquad &\text{if } f_\mathrm{bool}(x) = 1\,,\\
  0\,,\qquad &\text{if } f_\mathrm{bool}(x) = 0\,.\\
 \end{cases}
\end{equation}
In this case, it can be seen from \eqref{eq:bool_oracle}, \eqref{eq:gen_oracle}, and \eqref{eq:inv_gen_oracle} that the oracle $U_\varphi$ and its inverse $U^\dagger_\varphi$ both reduce to a boolean oracle as follows:
\begin{equation}
 U_{\varphi_\mathrm{bool}} = U^\dagger_{\varphi_\mathrm{bool}} = \hat{U}_{\!f_\mathrm{bool}}\,.
\end{equation}
Congruently, the task of amplifying (the amplitude of) the states with lower values of $\cos(\varphi)$ aligns with the task of amplifying the winning states $\Ket{x}$ with $f_\mathrm{bool}(x) = 1$.

\subsection{Goal of the Quantum Mean Estimation Algorithm}
Given a generic unitary operator $U$ and a state $\Ket{\psi_0}$, the goal of the quantum mean estimation algorithm introduced in this paper is to estimate the quantity
\begin{equation}
 \Braket{\psi_0|U|\psi_0}\,.
\end{equation}
In this paper, this task will be phrased in terms of the oracle $U_\varphi$ as estimating the expectation of the eigenvalue $e^{i\varphi(x)}$, for a state $\Ket{x}$ chosen randomly by making a measurement on the superposition state $\Ket{\psi_0}$. The connection between the two tasks can be seen, using \eqref{eq:gen_oracle_def}, as follows:
\begin{equation}
 \Braket{\psi_0|U_\varphi|\psi_0} = \sum_{x=0}^{N-1} \Bra{\big.\psi_0}e^{i\varphi(x)}\Ket{\big.x}\Braket{\big.x|\psi_0} = \sum_{x=0}^{N-1} \big|\Braket{x|\psi_0}\big|^2\,e^{i\varphi(x)}\,.
\end{equation}
Here, $\big|\Braket{x|\psi_0}\big|^2$ is the probability for a measurement on $\Ket{\psi_0}$ to yield $x$. The only difference between
\begin{itemize}
 \item estimating $\Braket{\psi_0|U_\varphi|\psi_0}$ for an oracle $U_\varphi$ of the form in \eqref{eq:gen_oracle_def}, and
 \item estimating $\Braket{\psi_0|U|\psi_0}$ for a generic unitary operator $U$
\end{itemize}
is that $\{\Ket{0},\dots,\Ket{N-1}\}$ is known, beforehand, to be an eigenbasis of $U_\varphi$. On the other hand, the eigenstates of a generic unitary operator $U$ may be a priori unknown. However, as we will see, the mean estimation algorithm of this paper does not use the knowledge of the eigenstates, and hence is applicable for generic unitary operators $U$ as well.

As mentioned before, the mean estimation algorithm of this paper is a generalization of the amplitude estimation algorithm of Ref.~\cite{QAAAE}. To see the connection between the respective tasks of these algorithms, note that the eigenvalues of a boolean oracle are either $+1$ or $-1$, and the expectation of the eigenvalue under $\Ket{\psi_0}$ is directly related to the probability of a measurement yielding a winning state with eigenvalue $-1$. This probability is precisely the quantity estimated by the amplitude estimation algorithm.
\vskip 5mm

The rest of the paper is organized as follows: The non-boolean amplitude amplification algorithm is described in \sref{sec:nbaaa} and analyzed in \sref{sec:anal}. \Sref{sec:mean} contains the description and analysis of the quantum mean estimation algorithm. Both these algorithms are demonstrated using a toy example in \sref{sec:toy}. Slightly modified versions of the algorithms are provided in \sref{sec:no_ancilla}. The tasks performed by these modified algorithms are related to, but different from, the tasks of the algorithms introduced in \sref{sec:nbaaa} and \sref{sec:mean}. Finally, the findings of this study and summarized and contextualized briefly in \sref{sec:summary}.

\section{Non-Boolean Amplitude Amplification Algorithm} \label{sec:nbaaa}

\subsection{Setup and Notation}
In addition to the quantum system (or qubits) that serve as inputs to the quantum oracle, the non-boolean amplitude amplification algorithm introduced in this section will use one extra ancilla qubit. For concreteness, let the quantum system used in the algorithm consist of two quantum registers. The first register contains the lone ancilla qubit, and the second register will be acted upon by the quantum oracle. 

The notations $\Ket{a}\otimes\Ket{b}$ and $\Ket{a,b}$ will both refer to the state where the two registers are unentangled, with the first register in state $\Ket{a}$ and the second register in state $\Ket{b}$. The tensor product notation $\otimes$ will also be used to combine operators that act on the individual registers into operators that simultaneously act on both registers. Such two-register operators will be represented by boldface symbols, e.g., $\Udif$, $\Uphi$, $\I$. Likewise, boldface symbols will be used to represent the states of the two-register system in the bra--ket notation, e.g., $\Ket{\Psi_0}$. Throughout this paper, any state written in the bra--ket notation, e.g., $\Ket{\psi}$, will be unit normalized, i.e., normalized to 1. The dagger notation ($\dagger$) will be used to denote the Hermitian conjugate of an operator, which is also the inverse for a unitary operator.

Throughout this paper, unless otherwise specified, $\{\Ket{0}, \Ket{1}, \dots, \Ket{N-1}\}$ will be used as the basis for (the state space of) the second register. Any measurement of the second register will refer to measurement in this basis. Likewise, unless otherwise specified,
\begin{equation}
 \Big\{\Ket{0,0}, \Ket{0,1}, \dots, \Ket{0,N-1}\Big\}~\cup~\Big\{\Ket{1,0}, \Ket{1,1}, \dots, \Ket{1,N-1}\Big\}
\end{equation}
will be used as the basis for the two-register system.

Let $\Ket{\psi_0}$ be the initial state of the second register from which the amplification process is to begin. Let $A_0$ be the unitary operator that changes the state of the second register from $\Ket{0}$ to $\Ket{\psi_0}$.\footnote{It is implicitly assumed here that there exists a state $\Ket{0}$ which is simultaneously an eigenstate of $U_\varphi$, as well as a special, easy-to-prepare state of the second register. The algorithms of this paper can be modified to work even without this assumption---it is made only for notational convenience.}
\begin{equation} \label{eq:psi0}
 \Ket{\psi_0} \equiv A_0\Ket{0} \equiv \sum_{x=0}^{N-1} a_0(x) \Ket{x}\,,\qquad \text{such that }\sum_{x=0}^{N-1} \big|a_0(x)\big|^2 = 1\,, 
\end{equation}
where $a_0(x)$ is the initial amplitude of the basis state $\Ket{x}$.

As we will see shortly, the algorithm introduced in this section will initialize the ancilla in the $\Ket{+}$ state given by
\begin{equation}
 \Ket{+} = \frac{\Ket{0}+\Ket{1}}{\sqrt{2}}\,.
\end{equation}
Anticipating this, let the two-register state $\Ket{\Psi_0}$ be defined as
\begin{subequations}\label{eq:Psi0}
\begin{align}
 \Ket{\Psi_0} &\equiv \Ket{+,\psi_0} = \frac{\Ket{0,\psi_0} + \Ket{1,\psi_0}}{\sqrt{2}}\\
 &= \frac{1}{\sqrt{2}}\sum_{x=0}^{N-1} a_0(x)\Big[\Ket{0, x} + \Ket{1, x}\Big]\,.
\end{align}
\end{subequations}

\subsection{Required Unitary Operations}
The following unitary operations will be used in the generalized amplitude amplification algorithm of this paper:

\subsubsection{Selective Phase-Flip Operator}
\begin{figure}[t]
 \centering
 \begin{tikzpicture}
  \node[scale=1.0] {
   \begin{quantikz}
    \lstick{$\displaystyle\genfrac{}{}{0pt}{}{\text{register 1}}{\text{(ancilla)}}$} & \qw & \gate{H}\gategroup[wires=2,steps=3,style={dashed,rounded corners,inner xsep=5pt,inner ysep=5pt,outer ysep=5pt},background]{Circuit for \,$\Udif$} & \gate[wires=2][10em]{2\Ket{0,0}\Bra{0,0} - \I} & \gate{H} & \qw & \qw \\[1em]
    \lstick{register 2} & \qwmult & \gate{A_0^\dagger} \qwmult & \qwmult & \gate{A_0} \qwmult & \qwmult & \qwmult
   \end{quantikz}
  };
 \end{tikzpicture}
 \caption{Quantum circuit for an implementation of $\Udif$, based on \eqref{eq:udif_circuit}.}
 \label{fig:udif}
\end{figure}
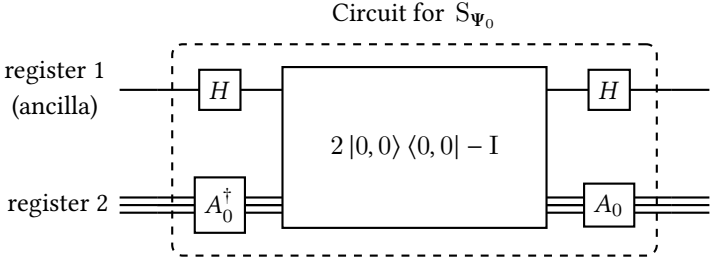

Let the two-register unitary operator $\Udif$ be defined as
\begin{subequations}\label{eq:udif}
\begin{align}
 \Udif &\equiv 2 \Ket{\Psi_0}\Bra{\Psi_0} - \I \\
 &= 2\Ket{+,\psi_0}\Bra{+, \psi_0} - \I\,,
\end{align}
\end{subequations}
where $\I$ is the two-register identity operator. $\Udif$ leaves the state $\Ket{\Psi_0}$ unchanged and flips the phase of any state orthogonal to $\Ket{\Psi_0}$. $\Udif$ is simply the two-register generalization of $S_{\psi_0}$ used in the boolean amplitude amplification algorithm. From \eqref{eq:psi0} and \eqref{eq:Psi0}, it can be seen that
\begin{equation}
 \Ket{\Psi_0} = \Big[H\otimes A_0\Big] \Ket{0,0}\,,
\end{equation}
where $H$ is the Hadamard transform. This allows $\Udif$ to be expressed as
\begin{equation}
 \Udif = \Big[H\otimes A_0\Big]~\Big[2 \Ket{0,0}\Bra{0,0} - \I\Big]~\Big[H\otimes A^\dagger_0\Big]\,. \label{eq:udif_circuit}
\end{equation}
This expression leads to an implementation of $\Udif$, as depicted in \fref{fig:udif}, provided one has access to the quantum circuits that implement $A_0$ and $A_0^\dagger$.

\subsubsection{Conditional Oracle Calls}
Let the two-register unitary operator $\Uphi$ be defined as
\begin{equation}
 \Uphi \equiv \Ket{0}\Bra{0} \otimes U_{\varphi} + \Ket{1}\Bra{1} \otimes U^\dagger_{\varphi}\,. \label{eq:Uphi}
\end{equation}
Its action on the basis states of the two-register system is given by
\begin{alignat}{2}
 \Uphi \Ket{0, x} &= e^{+i\varphi(x)} &&\Ket{0, x}\,,\\
 \Uphi \Ket{1, x} &= e^{-i\varphi(x)} &&\Ket{1, x}\,.
\end{alignat}
If the ancilla is in state $\Ket{0}$, $\Uphi$ acts $U_\varphi$ on the second register. On the other hand, if the ancilla is in state $\Ket{1}$, it acts $U^\dagger_\varphi$ on the second register. The inverse of $\Uphi$ is given by
\begin{equation}
 \Uphi^\dagger = \Ket{0}\Bra{0} \otimes U^\dagger_{\varphi} + \Ket{1}\Bra{1} \otimes U_{\varphi}\,, \label{eq:Uphidagger}
\end{equation}
and the action of $\Uphi^\dagger$ on the basis states is given by
\begin{alignat}{2}
 \Uphi^\dagger \Ket{0, x} &= e^{-i\varphi(x)} &&\Ket{0, x}\,,\\
 \Uphi^\dagger \Ket{1, x} &= e^{+i\varphi(x)} &&\Ket{1, x}\,.
\end{alignat}
The amplitude amplification algorithm for non-boolean functions will involve calls to both $\Uphi$ and $\Uphi^\dagger$. \Fref{fig:Uphi} and \fref{fig:Uphidagger} depict implementations of $\Uphi$ and $\Uphi^\dagger$ using a) the bit-flip (or Pauli-X) gate $X$, and b) controlled $U_\varphi$ and $U^\dagger_\varphi$ operations, with the ancilla serving as the control qubit.
\begin{figure}[t]
 \centering
 \begin{tikzpicture}
  \node[scale=1.0] {
   \begin{quantikz}
    \lstick{$\displaystyle\genfrac{}{}{0pt}{}{\text{register 1}}{\text{(ancilla)}}$} & \qw & \gate{X}\gategroup[wires=2,steps=4,style={dashed,rounded corners,inner xsep=5pt,inner ysep=5pt,outer ysep=5pt},background]{Circuit for \,$\Uphi$} & \ctrl{1} & \gate{X} & \ctrl{1} & \qw & \qw \\[1em]
    \lstick{register 2} & \qwmult & \qwmult & \gate{U_\varphi} \qwmult & \qwmult & \gate{U^\dagger_\varphi} \qwmult & \qwmult & \qwmult
   \end{quantikz}
  };
 \end{tikzpicture}
 \caption{Quantum circuit for an implementation of $\Uphi$ using controlled calls to the $U_\varphi$ and $U^\dagger_\varphi$ oracles.}
 \label{fig:Uphi}
 \vskip 5mm
 \begin{tikzpicture}
  \node[scale=1.0] {
   \begin{quantikz}
    \lstick{$\displaystyle\genfrac{}{}{0pt}{}{\text{register 1}}{\text{(ancilla)}}$} & \qw & \ctrl{1}\gategroup[wires=2,steps=4,style={dashed,rounded corners,inner xsep=5pt,inner ysep=5pt,outer ysep=5pt},background]{Circuit for \,$\Uphi^\dagger$} & \gate{X} & \ctrl{1} & \gate{X} & \qw & \qw \\[1em]
    \lstick{register 2} & \qwmult & \gate{U_\varphi} \qwmult & \qwmult & \gate{U^\dagger_\varphi} \qwmult & \qwmult & \qwmult & \qwmult
   \end{quantikz}
  };
 \end{tikzpicture}
 \caption{Quantum circuit for an implementation of $\Uphi^\dagger$ using controlled calls to the $U_\varphi$ and $U^\dagger_\varphi$ oracles.}
 \label{fig:Uphidagger}
\end{figure}


\subsection{Algorithm Description} \label{subsec:alg_desc}
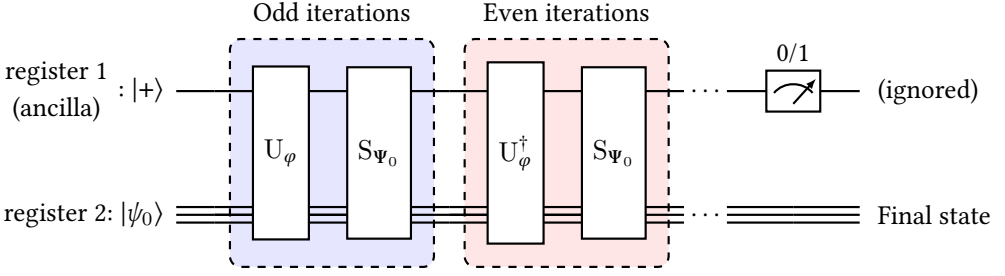
\begin{figure}[t]
 \centering
 \resizebox{\textwidth}{!}{
  \begin{tikzpicture}
   \node[scale=1.0] {
    \begin{quantikz}
     \lstick{$\displaystyle\genfrac{}{}{0pt}{}{\text{register 1}}{\text{(ancilla)}}$ : $\Ket{+}$} & \qw & \gate[wires=2]{\Uphi} \gategroup[wires=2,steps=2,style={dashed,rounded corners,fill=blue!10,inner xsep=5pt,inner ysep=5pt,outer ysep=3pt},background]{Odd iterations} & \gate[wires=2]{\Udif} & \qw & \gate[wires=2]{\Uphi^\dagger} \gategroup[wires=2,steps=2,style={dashed,rounded corners,fill=red!10,inner xsep=5pt,inner ysep=5pt,outer ysep=3pt},background]{Even iterations} & \gate[wires=2]{\Udif} & ~\ldots\qw~ & \meter{0/1} & \qw \rstick{~(ignored)} \\[1em]
     \lstick{register 2: $\Ket{\psi_0}$} & \qwmult & \qwmult & \qwmult & \qwmult & \qwmult & \qwmult & ~\ldots\qwmult~ & \qwmult & \qwmult \rstick{~Final state}
    \end{quantikz}
   };
  \end{tikzpicture}
 }
 \caption{Quantum circuit for the non-boolean amplitude amplification algorithm.}
 \label{fig:algorithm}
\end{figure}

The amplitude amplification algorithm for non-boolean functions is iterative and consists of the following steps:
\begin{enumerate}
 \item Initialize the two-register system in the $\Ket{\Psi_0}$ state.
 \item Perform $K$ iterations: During the odd iterations, apply the operation $\Udif\,\Uphi$ on the system. During the even iterations, apply $\Udif\,\Uphi^\dagger$ on the system.
 \item After the $K$ iterations, measure the ancilla (first register) in the 0/1 basis.
\end{enumerate}
Up to a certain number of iterations, the iterative steps are designed to amplify the amplitude of the basis states $\Ket{0,x}$ and $\Ket{1,x}$ with lower values of $\cos\!\big(\varphi(x)\big)$. 
The measurement of the ancilla at the end of the algorithm is performed simply to ensure that the two registers will not be entangled in the final state of the system. The quantum circuit and the pseudocode for the algorithm are shown in \fref{fig:algorithm} and \alref{alg:algorithm}, respectively. For pedagogical reasons, the specification of $K$, the number of iterations to perform, has been deferred to \sref{sec:anal}, which contains an analysis of the algorithm.
\begin{algorithm}[t]
 \begin{algorithmic}[1]
  \STATE \textbf{initialize} $\Ket{\Psi} := \Ket{\Psi_0}$
  \FOR{$k := 1$ \textbf{to} $K$}
   \IF{$k$ is odd}
    \STATE \textbf{update} $\Ket{\Psi} := \Udif\,\Uphi \Ket{\Psi}$
   \ELSE
    \STATE \textbf{update} $\Ket{\Psi} := \Udif\,\Uphi^\dagger\Ket{\Psi}$
   \ENDIF
  \ENDFOR
  \STATE Measure the ancilla in the 0/1 basis.
 \end{algorithmic}
 \caption{Non-boolean amplitude amplification algorithm of \sref{sec:nbaaa}.} \label{alg:algorithm}
\end{algorithm}

\subsection{Connection to the Boolean Amplitude Amplification Algorithm}
From \eqref{eq:Uphi} and \eqref{eq:Uphidagger}, it can be seen that for the boolean oracle case given by $\varphi(x) = \pi f_\mathrm{bool}(x)$, $\Uphi$ and $\Uphi^\dagger$ both reduce to $I\otimes \hat{U}_{\!f_\mathrm{bool}}$, where $\hat{U}_{\!f_\mathrm{bool}}$ is the oracle used in the boolean amplitude amplification algorithm. Furthermore, if the first register is in the $\Ket{+}$ state, from \eqref{eq:S_onereg} and \eqref{eq:udif}, the action of $\Udif$ is given by
\begin{equation}
 \Udif \Big[\Ket{+} \otimes \Ket{\psi}\Big] = \Ket{+} \otimes \Big[S_{\psi_0} \Ket{\psi}\Big]\,.
\end{equation}
Note that the first register is unaffected here. Thus, for the boolean oracle case, \alref{alg:algorithm} reduces to simply acting $S_{\psi_0}\,\hat{U}_{\!f_\mathrm{bool}}$ on the second register during each iteration---the ancilla qubit remains untouched and unentangled from the second register. In this way, the algorithm presented in this section is a generalization of the boolean amplitude amplification algorithm described in \sref{sec:intro}.

The two key differences of the generalized algorithm from the boolean one, apart from the usage of a non-boolean oracle, are
\begin{enumerate}
 \item The addition of the ancilla, which doubles the dimension of the state space of the system, and
 \item Alternating between using $\Uphi$ and $\Uphi^\dagger$ during the odd and even iterations.
\end{enumerate}
The motivation for these modifications will be provided in \sref{subsec:first_iter} and \sref{subsec:subseq_iters}, respectively.


\section{Analysis of the Non-Boolean Amplitude Amplification Algorithm} \label{sec:anal}
Let $\Ket{\Psi_k}$ be the state of the two-register system after $k = 0,1,\dots,K$ iterations of the amplitude amplification algorithm (but before the measurement of the ancilla). For $k>0$, $\Ket{\Psi_k}$ can be recursively written as
\begin{equation} \label{eq:Psik_recur}
 \Ket{\Psi_k} \equiv \begin{cases}
  \Udif\,\Uphi \Ket{\Psi_{k-1}},\qquad &\text{if } k \text{ is odd}\,,\\[2mm]
  \Udif\,\Uphi^\dagger \Ket{\Psi_{k-1}},\qquad &\text{if } k \text{ is even}\,.
 \end{cases}
\end{equation}
Let $\tilde{a}_k(0, x)$ and $\tilde{a}_k(1, x)$ be the normalized amplitudes of the basis states $\Ket{0,x}$ and $\Ket{1,x}$, respectively, in the superposition $\Ket{\Psi_k}$.
\begin{equation}
 \Ket{\Psi_k} \equiv \sum_{x=0}^N \Big[\tilde{a}_k(0, x)\Ket{0,x} + \tilde{a}_k(1, x)\Ket{1,x}\Big]\,.
\end{equation}
In the initial state $\Ket{\Psi_0}$, the amplitudes $\tilde{a}_0(0, x)$ and $\tilde{a}_0(1, x)$ are both given, from \eqref{eq:Psi0}, by \begin{equation}
 \tilde{a}_0(0, x) = \tilde{a}_0(1, x) = \frac{a_0(x)}{\sqrt{2}}\,.
\end{equation}

Let the parameter $\theta\in[0, \pi]$ be implicitly defined by
\begin{equation}
 \cos(\theta) \equiv \sum_{x=0}^{N-1} \big|a_0(x)\big|^2\,\cos\!\big(\varphi(x)\big)\,. \label{eq:costheta}
\end{equation}
$\cos(\theta)$ is the expected value of $\cos\!\big(\varphi(x)\big)$ over bitstrings $x$ sampled by measuring the state $\Ket{\psi_0}$.

Let the two-register states $\Ket{\Alpha}$ and $\Ket{\Beta}$ be defined as
\begin{align}
 \Ket{\Alpha} &\equiv \Uphi \Ket{\Psi_0}\,, \label{eq:alpha} \\
 \Ket{\Beta} &\equiv \Uphi^\dagger \Ket{\Psi_0}\,. \label{eq:beta}
\end{align}
They will be used to track the evolution of the system through the iterative steps of the algorithm. Using \eqref{eq:Psi0}, \eqref{eq:Uphi}, and \eqref{eq:Uphidagger}, $\Ket{\Alpha}$ and $\Ket{\Beta}$ can be written as
\begin{align}
 \Ket{\Alpha} &= \frac{1}{\sqrt{2}}\sum_{x=0}^{N-1}a_0(x)\Big[e^{i\varphi(x)}\Ket{0, x} + e^{-i\varphi(x)}\Ket{1, x}\Big]\,, \label{eq:alpha_expand}\\
 \Ket{\Beta} &= \frac{1}{\sqrt{2}}\sum_{x=0}^{N-1}a_0(x)\Big[e^{-i\varphi(x)}\Ket{0, x} + e^{i\varphi(x)}\Ket{1, x}\Big]\,. \label{eq:beta_expand}
\end{align}
Note that $\theta$, $\Ket{\Alpha}$, and $\Ket{\Beta}$ are all implicitly dependent on the function $\varphi$ and the initial state $\Ket{\psi_0}$. For notational convenience, these dependencies are not explicitly indicated. 



\subsection{The First Iteration} \label{subsec:first_iter}
After one iterative step, the system will be in state $\Ket{\Psi_1}$ given by
\begin{equation}
 \Ket{\Psi_1} = \Udif\,\Uphi \Ket{\Psi_0}\,.
\end{equation}
Using \eqref{eq:udif} and \eqref{eq:alpha}, this can be written as
\begin{equation}
 \Ket{\Psi_1} = \Udif \Ket{\Alpha} = 2\Braket{\Psi_0|\Alpha} \Ket{\Psi_0} - \Ket{\Alpha}\,. \label{eq:Psi1}
\end{equation}
From \eqref{eq:Psi0}, \eqref{eq:costheta}, and \eqref{eq:alpha_expand},
\begin{subequations} \label{eq:Psi0_alpha}
\begin{align}
 \Braket{\Psi_0|\Alpha} &= \frac{1}{2}\sum_{x=0}^{N-1}~ \big|a_0(x)\big|^2 ~\Big[e^{i\varphi(x)} \Braket{0,x|0,x} + e^{-i\varphi(x)} \Braket{1,x|1,x}\Big]\\
 &= \frac{1}{2}\sum_{x=0}^{N-1}~\big|a_0(x)\big|^2 ~\Big[e^{i\varphi(x)} + e^{-i\varphi(x)}\Big]\\
 &= \frac{1}{2}\sum_{x=0}^{N-1}~ \big|a_0(x)\big|^2~~2\cos\!\big(\varphi(x)\big) = \cos(\theta)\,.
\end{align}
\end{subequations}
Note that $\Braket{\Psi_0|\Alpha}$ is real-valued. This fact is crucial to the functioning of the algorithm, as we will see later in this subsection. The motivation behind adding an ancilla qubit, effectively doubling the number of basis states, is precisely to make $\Braket{\Psi_0|\Alpha}$ real-valued.

From \eqref{eq:Psi1} and \eqref{eq:Psi0_alpha}, $\Ket{\Psi_1}$ can be written as
\begin{equation}
 \Ket{\Psi_1} = 2\cos(\theta)\Ket{\Psi_0} - \Ket{\Alpha}\,.
\end{equation}
From \eqref{eq:Psi0} and \eqref{eq:alpha_expand}, the amplitude $\tilde{a}_1(0, x)$ of the basis state $\Ket{0,x}$ in the superposition $\Ket{\Psi_1}$ can be written as
\begin{equation}
 \tilde{a}_1(0, x) = \Braket{0,x|\Psi_1} = \frac{a_0(x)}{\sqrt{2}} \Big[2\cos(\theta) - e^{i\varphi(x)}\Big]\,. \label{eq:oneiter0}
\end{equation}
Likewise, the amplitude $\tilde{a}_1(1, x)$ of the basis state $\Ket{1,x}$ can be written as
\begin{equation}
 \tilde{a}_1(1, x) = \Braket{1,x|\Psi_1} = \frac{a_0(x)}{\sqrt{2}} \Big[2\cos(\theta) - e^{-i\varphi(x)}\Big]\,. \label{eq:oneiter1}
\end{equation}
From \eqref{eq:oneiter0} and \eqref{eq:oneiter1}, it can be seen that after one iterative step, the amplitudes of $\Ket{0,x}$ and $\Ket{1,x}$ have acquired factors of $\left[2\cos(\theta) - e^{i\varphi(x)}\right]$ and $\left[2\cos(\theta) - e^{-i\varphi(x)}\right]$, respectively. Now,
\begin{equation}
 \frac{\big|\tilde{a}_1(0, x)\big|^2}{\big|\tilde{a}_0(0, x)\big|^2} = \frac{\big|\tilde{a}_1(1, x)\big|^2}{\big|\tilde{a}_0(1, x)\big|^2} = 4\cos^2(\theta) + 1 - 4\cos(\theta)\cos\!\big(\varphi(x)\big)\,.
\end{equation}
This shows that if $\cos(\theta)$ is positive, the magnitude of the ``amplitude amplification factor'' after one iteration is monotonically decreasing in $\cos(\varphi)$. This preferential amplification of states with lower values of $\cos(\varphi)$ is precisely what the algorithm set out to do.

Note that this monotonicity property relies crucially on $\Braket{\Psi_0|\Alpha}$ being real-valued in \eqref{eq:Psi1}. If $\Braket{\Psi_0|\Alpha}$ is complex, with a phase $\delta \notin \{0,\pi\}$, then the amplification will be monotonic in $\cos(\varphi - \delta)$, which does not meet the present goal of the algorithm. The case where $\Braket{\Psi_0|\Alpha}$ is not real-valued is explored further in \sref{sec:no_ancilla}.

\subsection{Identities to Track the Subsequent Iterations} \label{subsec:subseq_iters}
Equations \eqref{eq:Psi0}, \eqref{eq:udif}, \eqref{eq:Uphi}, \eqref{eq:Uphidagger}, \eqref{eq:alpha}, and \eqref{eq:beta} can be used to derive the following identities, which capture the actions of the operators $\Udif$, $\Uphi$, and $\Uphi^\dagger$ on the states $\Ket{\Psi_0}$, $\Ket{\Alpha}$, and $\Ket{\Beta}$:
\begin{alignat}{2}
 \Udif &\Ket{\Psi_0} &&= \Ket{\Psi_0}\,,\\
 \Udif &\Ket{\Alpha} &&= 2\cos(\theta) \Ket{\Psi_0} - \Ket{\Alpha}\,,\\
 \Udif &\Ket{\Beta} &&= 2\cos(\theta) \Ket{\Psi_0} - \Ket{\Beta}\,,\\
 \Uphi &\Ket{\Psi_0} &&= \Ket{\Alpha}\,,\\
 \Uphi &\Ket{\Alpha} &&= \frac{1}{\sqrt{2}}\sum_{x=0}^{N-1}a_0(x) \Big[e^{2i\varphi(x)}\Ket{0,x} + e^{-2i\varphi(x)}\Ket{1,x}\Big]\,,\label{eq:Uphi_bad}\\
 \Uphi &\Ket{\Beta} &&= \Ket{\Psi_0}\,,\\
 \Uphi^\dagger &\Ket{\Psi_0} &&= \Ket{\Beta}\,,\\
 \Uphi^\dagger &\Ket{\Alpha} &&= \Ket{\Psi_0}\,,\\
 \Uphi^\dagger &\Ket{\Beta} &&= \frac{1}{\sqrt{2}}\sum_{x=0}^{N-1}a_0(x) \Big[e^{-2i\varphi(x)}\Ket{0,x} + e^{2i\varphi(x)}\Ket{1,x}\Big]\,\label{eq:Uphidagger_bad}.
\end{alignat}
We can see that the subspace spanned by the states $\Ket{\Psi_0}$, $\Ket{\Alpha}$, and $\Ket{\Beta}$ is \emph{almost} stable under the action of $\Udif$, $\Uphi$, and $\Uphi^\dagger$. Only the actions of $\Uphi$ on $\Ket{\Alpha}$ and $\Uphi^\dagger$ on $\Ket{\Beta}$ can take the state of the system out of this subspace. The motivation behind alternating between using $\Uphi$ and $\Uphi^\dagger$ during the odd and the even iterations is to keep the state of the system within the subspace spanned by $\Ket{\Psi_0}$, $\Ket{\Alpha}$, and $\Ket{\Beta}$.

\begin{figure}[t]
 \centering
 \tikzset{curve/.style={decorate, decoration={bent,amplitude=.7em,aspect=.3}}}
 \tikzstyle{block} = [rectangle, draw, text width=10em, text centered, rounded corners, minimum height=7em]
 \tikzstyle{arrow} = [draw, arrows = {-Latex[length=.5em 3 0]}]
 \begin{tikzpicture}[node distance = 20em, auto]
  \node [block,fill=red!7] (betaspace) {Subspace spanned by $\Ket{\Psi_0}$ and $\Ket{\Beta}$};
  \node [block,fill=blue!7, right of=betaspace] (alphaspace) {Subspace spanned by $\Ket{\Psi_0}$ and $\Ket{\Alpha}$};
  
  \path [arrow, transform canvas={yshift=2em}, draw=blue, curve] (betaspace) -- node [above,yshift=.5em] {\textcolor{blue}{odd iterations}} (alphaspace);
  \path [arrow, transform canvas={yshift=-2em}, draw=red, curve] (alphaspace) -- node [below,yshift=-.5em] {\textcolor{red}{even iterations}} (betaspace);
 \end{tikzpicture}
 \caption{Illustration depicting the evolution of the state of two-register system through the iterations of the non-boolean amplitude amplification algorithm.}
 \label{fig:odd_even_subspaces}
\end{figure}
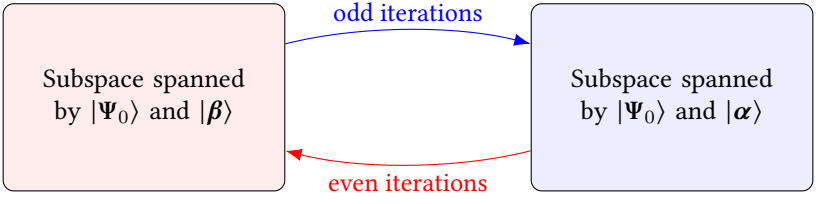

From the identities above, we can write the following expressions capturing the relevant actions of the odd iteration operator $\Udif\,\Uphi$ and the even iteration operator $\Udif\,\Uphi^\dagger$:
\begin{alignat}{2}
 \Udif\,\Uphi &\Ket{\Psi_0} &&= 2\cos(\theta) \Ket{\Psi_0} - \Ket{\Alpha}\,, \label{eq:iter_iden_start}\\
 \Udif\,\Uphi &\Ket{\Beta} &&= \Ket{\Psi_0}\,, \label{eq:iter_iden_2}\\
 \Udif\,\Uphi^\dagger &\Ket{\Psi_0} &&= 2\cos(\theta) \Ket{\Psi_0} - \Ket{\Beta}\,,\\
 \Udif\,\Uphi^\dagger &\Ket{\Alpha} &&= \Ket{\Psi_0}\,. \label{eq:iter_iden_end}
\end{alignat}
From these expressions, it can be seen that the odd iteration operator $\Udif\,\Uphi$ maps any state in the space spanned by $\Ket{\Psi_0}$ and $\Ket{\Beta}$ to a state in the space spanned by $\Ket{\Psi_0}$ and $\Ket{\Alpha}$. Conversely, the even iteration operator $\Udif\,\Uphi^\dagger$ maps any state in the space spanned by $\Ket{\Psi_0}$ and $\Ket{\Alpha}$ to a state in the space spanned by $\Ket{\Psi_0}$ and $\Ket{\Beta}$. Since the algorithm begins with the system initialized in the state $\Ket{\Psi_0}$, it can be seen that the state of the system oscillates between the two subspaces during the odd and even iterations, as depicted in \fref{fig:odd_even_subspaces}.

\subsection{State After \texorpdfstring{$k$}{k} Iterations}
Using \eqref{eq:Psik_recur} and (\ref{eq:iter_iden_start}--\ref{eq:iter_iden_end}), the state $\Ket{\Psi_k}$ of the two-register system after $k\geq 0$ iterations can be written, in matrix multiplication notation, as
\begin{equation} \label{eq:Psik_matrix}
 \Ket{\Psi_k} = \begin{cases}
  \begin{matrix}
   \begin{bmatrix}
    \Ket{\Psi_0} ~&~ \Ket{\Alpha}
   \end{bmatrix}\\
   \mbox{}
  \end{matrix}
  \begin{bmatrix}
   2\cos(\theta) ~&~ 1\\[.5em]
   -1 ~&~ 0
  \end{bmatrix}^k
  \begin{bmatrix}
   1\\[.5em]
   0
  \end{bmatrix},\qquad\qquad &\text{if } k \text{ is odd}\,,\\[2.5em]
  \begin{matrix}
   \begin{bmatrix}
    \Ket{\Psi_0} ~&~ \Ket{\Beta}
   \end{bmatrix}\\
   \mbox{}
  \end{matrix}
  \begin{bmatrix}
   2\cos(\theta) ~&~ 1\\[.5em]
   -1 ~&~ 0
  \end{bmatrix}^k
  \begin{bmatrix}
   1\\[.5em]
   0
  \end{bmatrix},\qquad\qquad &\text{if } k \text{ is even}\,.
 \end{cases}
\end{equation}
As shown in \aref{appendix:state_derivation}, \eqref{eq:Psik_matrix} can be simplified to
\begin{equation} \label{eq:Psik_full}
 \Ket{\Psi_k} = \begin{cases}
  \displaystyle\frac{1}{\sin(\theta)} \Big[\sin\!\big((k+1)\theta\big) \Ket{\Psi_0} -\sin(k\theta) \Ket{\Alpha}\Big]\,,\qquad\qquad &\text{if } k \text{ is odd}\,,\\[1.5em]
  \displaystyle\frac{1}{\sin(\theta)} \Big[\sin\!\big((k+1)\theta\big) \Ket{\Psi_0} -\sin(k\theta) \Ket{\Beta}\Big]\,,\qquad\qquad &\text{if } k \text{ is even}\,.
 \end{cases}
\end{equation}

\subsection{Basis State Amplitudes After \texorpdfstring{$k$}{k} Iterations} \label{subsec:state_ampl}
From \eqref{eq:Psi0}, \eqref{eq:alpha_expand}, \eqref{eq:beta_expand}, and \eqref{eq:Psik_full}, the amplitudes $\tilde{a}_k(0, x)$ of the basis states $\Ket{0,x}$ after $k\geq 0$ iterations can be written as
\begin{equation}
 \tilde{a}_k(0, x) = \Braket{0,x|\Psi_k} = \begin{cases}
  \displaystyle\frac{a_0(x)}{\sqrt{2}~\sin(\theta)} \Big[\sin\!\big((k+1)\theta\big) - \sin(k\theta)e^{i\varphi(x)}\Big]\,,\qquad &\text{if } k \text{ is odd}\,,\\[1.5em]
  \displaystyle\frac{a_0(x)}{\sqrt{2}~\sin(\theta)} \Big[\sin\!\big((k+1)\theta\big) - \sin(k\theta)e^{-i\varphi(x)}\Big]\,,\qquad &\text{if } k \text{ is even}\,.
 \end{cases}
\end{equation}
Similarly, the amplitudes $\tilde{a}_k(1, x)$ of the basis states $\Ket{1,x}$ can be written as
\begin{equation}
 \tilde{a}_k(1, x) = \Braket{1,x|\Psi_k} = \begin{cases}
  \displaystyle\frac{a_0(x)}{\sqrt{2}~\sin(\theta)} \Big[\sin\!\big((k+1)\theta\big) - \sin(k\theta)e^{-i\varphi(x)}\Big]\,,\qquad &\text{if } k \text{ is odd}\,,\\[1.5em]
  \displaystyle\frac{a_0(x)}{\sqrt{2}~\sin(\theta)} \Big[\sin\!\big((k+1)\theta\big) - \sin(k\theta)e^{i\varphi(x)}\Big]\,,\qquad &\text{if } k \text{ is even}\,.
 \end{cases}
\end{equation}
These expressions can be summarized, for $b\in\{0,1\}$, as
\begin{equation} \label{eq:tworeg_state_ampl}
 \tilde{a}_k(b, x) = \begin{cases}
  \displaystyle\frac{a_0(x)}{\sqrt{2}~\sin(\theta)} \Big[\sin\!\big((k+1)\theta\big) - \sin(k\theta)e^{i\varphi(x)}\Big]\,,\qquad &\text{if } k+b \text{ is odd}\,,\\[1.5em]
  \displaystyle\frac{a_0(x)}{\sqrt{2}~\sin(\theta)} \Big[\sin\!\big((k+1)\theta\big) - \sin(k\theta)e^{-i\varphi(x)}\Big]\,,\qquad &\text{if } k+b \text{ is even}\,.
 \end{cases}
\end{equation}
\subsubsection*{Amplitudes After Ancilla Measurement}

Note that the magnitudes of the amplitudes of the states $\Ket{0,x}$ and $\Ket{1,x}$ are equal, i.e.,
\begin{equation}
 \big|\tilde{a}_k(0, x)\big| = \big|\tilde{a}_k(1, x)\big|\,,
\end{equation}
for all $k\geq 0$ and $x\in\{0,1,\dots,N-1\}$. So, a measurement of the ancilla qubit in the first register after $K$ iterations will yield a value of either 0 or 1 with equal probability. Let $\Ket{\psi_{K,b}}$ be the normalized state of the second register after performing $K$ iterations, followed by a measurement of the ancilla qubit, which yields a value $b\in\{0,1\}$. $\Ket{\psi_{K,b}}$ can be written as
\begin{equation}
 \Ket{\psi_{K,b}} = \sum_{x=0}^{N-1} a_{K,b}(x) \Ket{x}\,,
\end{equation}
where $a_{K,b}(x)$ are the normalized amplitudes of the basis states of the second register (after performing $K$ iterations and the measurement of the ancilla). $a_{K,b}(x)$ is simply given by
\begin{equation}
 a_{K,b}(x) = \sqrt{2}~\tilde{a}_K(b, x)\,.
\end{equation}
Much of the following discussion holds a) regardless of whether or not a measurement is performed on the ancilla qubit after the $K$ iterations, and b) regardless of the value yielded by the ancilla measurement (if performed)---the primary goal of measuring the ancilla is simply to 
make the two registers unentangled from each other.

\subsection{Basis State Probabilities After \texorpdfstring{$K$}{K} Iterations} \label{subsec:state_prob}
Let $p_K(x)$ be the probability for a measurement of the second register after $K\geq 0$ iterations to yield $x$. It can be written in terms of the amplitudes in \sref{subsec:state_ampl} as
\begin{equation}
 p_K(x) = \left[\big|\tilde{a}_K(0, x)\big|^2 + \big|\tilde{a}_K(1, x)\big|^2\right] = \big|a_{K,0}(x)\big|^2 = \big|a_{K,1}(x)\big|^2\,. \label{eq:pk_equiv}
\end{equation}
This expression shows that the probability $p_K(x)$ depends neither on whether the ancilla was measured, nor on the result of the ancilla measurement (if performed). From \eqref{eq:tworeg_state_ampl}, $p_K(x)$ can be written as
\begin{align}
 p_K&(x) = \frac{p_0(x)}{\sin^2(\theta)}\Big|\sin\!\big((K+1)\theta\big) - \sin(K\theta)e^{i\varphi(x)}\Big|^2\\
 &= \frac{p_0(x)}{\sin^2(\theta)}\Big[\sin^2(K\theta) + \sin^2\!\big((K+1)\theta\big) - 2\,\sin(K\theta)\,\sin\!\big((K+1)\theta\big)\,\cos\!\big(\varphi(x)\big)\Big]\,. \label{eq:Pk_long}
\end{align}
It can be seen here that the probability amplification factor $p_K(x)/p_0(x)$ is monotonic in $\cos\!\big(\varphi(x)\big)$ for all $K\geq 0$. The following trigonometric identities, proved in \aref{appendix:trig_proofs}, help 
elucidate the $K$ dependence of this amplification factor:
\begin{align}
 \sin^2(C) + \sin^2(C+D) &= \sin^2(D) + 2\,\sin(C)\,\sin(C+D)\,\cos(D)\,, \label{eq:trig1}\\
 2\,\sin(C)\,\sin(C+D) &= \cos(D) - \cos(2C+D)\,. \label{eq:trig2}
\end{align}
Setting $C=K\theta$ and $D=\theta$, these identities can be used to rewrite \eqref{eq:Pk_long} as
\begin{equation}
 p_K(x) = p_0(x) \Big\{1 - \lambda_K \Big[\cos\!\big(\varphi(x)\big) - \cos(\theta)\Big]\Big\}\,, \label{eq:Pk_short}
\end{equation}
where the $K$-dependent factor $\lambda_K$ is given by
\begin{equation}
 \lambda_K \equiv \frac{2\,\sin(K\theta)\,\sin\!\big((K+1)\theta\big)}{\sin^2(\theta)} = \frac{\cos(\theta) - \cos\!\big((2K+1)\theta\big)}{\sin^2(\theta)}\,. \label{eq:lambda}
\end{equation}
For notational convenience, the fact that $\lambda_K$ depends on $\theta$ is not explicitly indicated. The result in \eqref{eq:Pk_short} can be summarized as follows:
\begin{itemize}
 \item Applying $K$ iterations of the non-boolean amplitude amplification algorithm changes the probability of measuring $x$ by a factor that is a linear function of $\cos\!\big(\varphi(x)\big)$. If the second register is initially in an equiprobable state, i.e., if $p_0(x)=\mathrm{constant}$, then the probability $p_K(x)$ after $K$ iterations is itself a linear function of $\cos\!\big(\varphi(x)\big)$.
 \item If $\cos\!\big(\varphi(x)\big) = \cos(\theta)$ for some $x$, the probability of a measurement of the second register yielding that $x$ is unaffected by the algorithm.
 \item The slope of the linear dependence is $-\lambda_K$.
 \begin{itemize}
  \item If $\lambda_K$ is positive, the states with $\cos(\varphi) < \cos(\theta)$ are amplified. Conversely, if $\lambda_K$ is negative, states with $\cos(\varphi) > \cos(\theta)$ are amplified.
  \item The magnitude of $\lambda_K$ controls the degree to which the preferential amplification has been performed.
 \end{itemize}
\end{itemize}
From \eqref{eq:lambda}, it can be seen that $\lambda_K$ is an oscillatory function of $K$, centered around $\cos(\theta)/\sin^2(\theta)$, with an amplitude of $1/\sin^2(\theta)$ and a period of $\pi/\theta$.
Recalling from \eqref{eq:costheta} that
\begin{equation}
 \sum_{x=0}^{N-1} p_0(x)\,\cos\!\big(\varphi(x)\big) = \cos(\theta) = \sum_{x=0}^{N-1}p_0(x)\,\cos(\theta)\,,
\end{equation}
one can verify that for any $K\geq 0$, the probabilities $p_K(x)$ from \eqref{eq:Pk_short} add up to 1.

From the definition of $\lambda_K$ in \eqref{eq:lambda}, it can be seen that for all $K$, $\lambda_K$ is bounded from above by $\lambda_\mathrm{optimal}$ defined as
\begin{equation} \label{eq:lambda_opt}
 \lambda_K \leq \lambda_\mathrm{optimal} \equiv \frac{\cos(\theta)+1}{\sin^2(\theta)} = \frac{1}{1-\cos(\theta)}\,.
\end{equation}
The $\lambda_K = \lambda_\mathrm{optimal}$ case represents the maximal preferential amplification of lower values of $\cos(\varphi)$ achievable by the algorithm. Let $p_\mathrm{optimal}(x)$ be the state probability function corresponding to $\lambda_K = \lambda_\mathrm{optimal}$. From \eqref{eq:Pk_short},
\begin{equation}
 p_\mathrm{optimal}(x) = p_0(x)\left[1 - \frac{\cos\!\big(\varphi(x)\big) - \cos(\theta)}{1 - \cos(\theta)}\right]\,.
\end{equation}
It is interesting to note that $p_\mathrm{optimal}(x) = 0$ for inputs $x$ with the highest possible value of $\cos\!\big(\varphi(x)\big)$, namely 1. In other words, $p_\mathrm{optimal}$ reaches the limit set by the non-negativity of probabilities, in the context of the algorithm under consideration.

\subsection{Number of Iterations to Perform} \label{subsec:optimalK}
In the description of the algorithm in \sref{sec:nbaaa}, the number of iterations $K$ to perform was left unspecified. Armed with \eqref{eq:lambda}, this aspect of the algorithm can now be tackled. Higher values of $\lambda_K$ are preferable for the purposes of this paper, namely to preferentially amplify \emph{lower} values of $\cos(\varphi)$. From \eqref{eq:lambda}, it can be seen that $\lambda_K$ is monotonically increasing for $K=0,1,\dots$ as long as $0 \leq (2K+1)\theta \leq \pi + \theta$ or, equivalently, for
\begin{equation}
 0 \leq K \leq \left\lfloor\frac{\pi}{2\theta}\right\rfloor\,,
\end{equation}
where $\lfloor v\rfloor$ denotes the floor of $v$. As with the boolean amplitude amplification algorithm of Ref.~\cite{QAAAE}, a good approach is to stop the algorithm just before the first iteration that, if performed, would cause value of $\lambda_K$ to decrease (from its value after the previous iteration). This leads to the choice $\tilde{K}$ for the number of iterations to perform, given by
\begin{equation} \label{eq:K_heur}
 \tilde{K} = \left\lfloor\frac{\pi}{2\theta}\right\rfloor\,.
\end{equation}
The corresponding value of $\lambda_K$ for $K=\tilde{K}$ is given by
\begin{equation} \label{eq:lambda_heur}
 \lambda_{\tilde{K}} = \frac{1}{\sin^2(\theta)}\Big[\cos(\theta) - \cos\!\left(\left\lfloor\frac{\pi}{2\theta}\right\rfloor 2\theta + \theta\right)\Big]\,.
\end{equation}
The choice $\tilde{K}$ in \eqref{eq:K_heur} for the number of iterations offers an amplification iff $\pi > 2\theta > 0$ or, equivalently, iff $0 < \cos(\theta) < 1$. At one of the extremes, namely $\theta=\pi/2$, we have $\lambda_{\tilde{K}} = 0$. The other extreme, namely $\cos(\theta)=1$, corresponds to every state $x$ with a non-zero amplitude in the initial state $\Ket{\psi_0}$ having $\cos\!\big(\varphi(x)\big)=1$; there is no scope for preferential amplification in this case.

From \eqref{eq:lambda_heur}, it can be seen that $\lambda_{\tilde{K}}$ exactly equals $\lambda_\mathrm{optimal}$ defined in \eqref{eq:lambda_opt} if $\pi/(2\theta)$ is a half-integer. In terms of $\theta$, this condition can be written as
\begin{equation}
 \theta\in\left\{\frac{\pi}{3}, \frac{\pi}{5}, \frac{\pi}{7}, \dots\text{(harmonic progression)}\right\}\quad \Rightarrow \quad \lambda_{\tilde{K}} = \lambda_\mathrm{optimal}\,.
\end{equation}
For generic values of $\theta$, from \eqref{eq:lambda_heur}, it can be seen that $\lambda_{\tilde{K}}$ satisfies
\begin{equation}
 \frac{2\cos(\theta)}{\sin^2(\theta)} \leq \lambda_{\tilde{K}} \leq \lambda_\mathrm{optimal} = \frac{\cos(\theta)+1}{\sin^2(\theta)}\,.
\end{equation}
This can be rewritten as
\begin{equation} \label{eq:lambda_bound}
 \lambda_\mathrm{optimal} \Big[1-\tan^2(\theta/2)\Big] \leq \lambda_{\tilde{K}} \leq \lambda_\mathrm{optimal}\,,
\end{equation}
using the following identity:
\begin{equation}
 \frac{2\cos(\theta)}{\cos(\theta) + 1} = \frac{2 \Big[\cos^2(\theta/2) - \sin^2(\theta/2)\Big]}{2\cos^2(\theta/2) - 1 + 1} = 1 - \tan^2(\theta/2)\,.
\end{equation}
From \eqref{eq:lambda_bound}, it can be seen that for small $\theta$, $\lambda_{\tilde{K}}$ is approximately equal to $\lambda_\mathrm{optimal}$, within an $\mathcal{O}(\theta^2)$ relative error.

\subsection{Mean and Higher Moments of \texorpdfstring{$\cos(\varphi)$}{cos(phi)} After \texorpdfstring{$K$}{K} Iterations} \label{subsec:moments}

Let $\mu^{(n)}_K$ be the $n$-th raw moment of $\cos\!\big(\varphi(x)\big)$ for a random value of $x$ sampled by measuring the second register after $K$ iterations.
\begin{equation}
 \mu^{(n)}_K \equiv \sum_{x=0}^{N-1} p_K(x)\,\cos^n\!\big(\varphi(x)\big)\,.
\end{equation}
Under this notation, $\mu^{(1)}_0$ is simply $\cos(\theta)$. From \eqref{eq:Pk_short}, we can write $\mu^{(n)}_K$ in terms of the initial moments ($K=0$) as
\begin{subequations} \label{eq:raw_moments}
\begin{align}
 \mu^{(n)}_K &= \sum_{x=0}^{N-1}~p_0(x)\,\cos^n\!\big(\varphi(x)\big)\,\Big[1 - \lambda_K \Big[\cos\!\big(\varphi(x)\big) - \cos(\theta)\Big]\Big]\\
 &= \mu^{(n)}_0 - \lambda_K \Big[\mu^{(n+1)}_0 - \mu^{(n)}_0 \mu^{(1)}_0\Big]\,.
\end{align}
\end{subequations}
In particular, let $\mu_K$ and $\sigma^2_K$ represent the expected value and variance, respectively, of $\cos\!\big(\varphi(x)\big)$ after $K$ iterations.
\begin{align}
 \mu_K &\equiv \mu^{(1)}_K\,,\\
 \sigma^2_K &\equiv \mu^{(2)}_K - \left[\mu^{(1)}_K\right]^2\,.
\end{align}
Now, the result in \eqref{eq:raw_moments} for $n=1$ can be written as
\begin{equation}
 \mu_K - \mu_0 = -\lambda_K \sigma^2_0\,.
\end{equation}
For $\lambda_K>0$, this equation captures the reduction in the expected value of $\cos\!\big(\varphi(x)\big)$ resulting from $K$ iterations of the algorithm.

\subsection{Cumulative Distribution Function of \texorpdfstring{$\cos(\varphi)$}{cos(phi(x))} After \texorpdfstring{$K$}{K} Iterations} \label{subsec:cumulative}
Let $F^{\cos}_K(y)$ be the probability that $\cos\!\big(\varphi(x)\big) \leq y$, for an $x$ sampled as per the probability distribution $p_K$. $F^{\cos}_K$ is the cumulative distribution function of $\cos(\varphi)$ for a measurement after $K$ iterations, and can be written as
\begin{align}
 F^{\cos}_K(y) &= \sum_{x=0}^{N-1}\Big[p_K(x)~~\mathbbm{1}_{[0,\infty)}\!\left[y - \cos\!\big(\varphi(x)\big)\right]\Big]\,,\\
 1-F^{\cos}_K(y) &= \sum_{x=0}^{N-1}\Big[~p_K(x)~~\Big(1-\mathbbm{1}_{[0,\infty)}\!\left[y - \cos\!\big(\varphi(x)\big)\right]\Big)~\Big]\,,
\end{align}
where $\mathbbm{1}_{[0,\infty)}$ is the Heaviside step function, which equals $0$ when its argument is negative and $1$ when its argument is non-negative. Using the expression for $p_K$ in \eqref{eq:Pk_short}, these can be written as
\begin{align}
 F^{\cos}_K(y) &= F^{\cos}_0(y) \Big(1+\lambda_K\,\mu_0\Big) - \lambda_K~\sum_{x=0}^{N-1}\Big\{p_0(x)\,\cos\!\big(\varphi(x)\big)~~\mathbbm{1}_{[0,\infty)}\!\left[y - \cos\!\big(\varphi(x)\big)\right]\Big\}\,, \label{eq:Fk_expand}\\
 \begin{split}
  1-F^{\cos}_K(y) &= \Big(1-F^{\cos}_0(y)\Big) \Big(1+\lambda_K\,\mu_0\Big) \\
  &\qquad\qquad- \lambda_K~\sum_{x=0}^{N-1}\Big\{p_0(x)\,\cos\!\big(\varphi(x)\big)~~\Big[1-\mathbbm{1}_{[0,\infty)}\!\left[y - \cos\!\big(\varphi(x)\big)\right]\Big]\Big\}\,. \label{eq:1minusFk_expand}
 \end{split}
\end{align}
Every $x$ that provides a non-zero contribution to the summation in \eqref{eq:Fk_expand} satisfies $\cos\!\big(\varphi(x)\big) \leq y$.  This fact can be used to write
\begin{equation}
 \lambda_K\geq 0 \quad \Rightarrow \qquad\quad F^{\cos}_K(y) \geq F^{\cos}_0(y) \Big(1+\lambda_K\big(\mu_0-y\big)\Big)\,. \label{eq:cum_ineq_1}
\end{equation}
Likewise, every $x$ that provides a non-zero contribution to the summation in \eqref{eq:1minusFk_expand} satisfies $\cos\!\big(\varphi(x)\big) > y$. This can be used to write
\begin{equation}
 \lambda_K\geq 0 \quad \Rightarrow \qquad\quad 1-F^{\cos}_K(y) \leq \Big(1-F^{\cos}_0(y)\Big) \Big(1+\lambda_K\big(\mu_0-y\big)\Big)\,. \label{eq:cum_ineq_2}
\end{equation}
The inequalities in \eqref{eq:cum_ineq_1} and \eqref{eq:cum_ineq_2} can be summarized as
\begin{equation}
 \lambda_K\geq 0 \quad \Rightarrow \qquad F^{\cos}_k(y) \geq F^{\cos}_0(y) + \lambda_K\,\max\left\{F^{\cos}_0(y)\big(\mu_0-y\big), \Big(1-F^{\cos}_0(y)\Big)\big(y-\mu_0\big)\right\}\,, \label{eq:cum_ineq}
\end{equation}
where the $\max$ function represents the maximum of its two arguments. This equation provides a lower bound on the probability that a measurement after $K$ iterations yields a state whose $\cos(\varphi)$ value is no higher than $y$. It may be possible to derive stronger bounds (or even the exact expression) for $F^{\cos}_K(y)$ if additional information is known about the initial distribution of $\cos(\varphi)$. For $y\leq \mu_0$, the first argument of the $\max$ function in \eqref{eq:cum_ineq} will be active, and for $y\geq \mu_0$, the second argument will be active.

For the $\lambda_K\leq 0$ case, it can similarly be shown that
\begin{equation}
 \lambda_K\leq 0 \quad \Rightarrow \qquad F^{\cos}_K(y) \leq F^{\cos}_0(y) + \lambda_K\,\min\left\{F^{\cos}_0(y)\big(\mu_0-y\big), \Big(1-F^{\cos}_0(y)\Big)\big(y-\mu_0\big)\right\}\,, \label{eq:cum_ineq_alt}
\end{equation}
where the $\min$ function represents the minimum of its two arguments.

\subsection{Boolean Oracle Case} \label{subsec:boolean_case}
The result in \sref{subsec:optimalK} for the heuristic choice for the number of iterations, namely $\tilde{K} = \lfloor\pi/(2\theta)\rfloor$, might be reminiscent of the analogous result for the boolean amplitude amplification algorithm in Ref.~\cite{QAAAE}, namely $\lfloor\pi/(4\theta_a)\rfloor$.
The similarity between the two results is not accidental. To see this, consider the parameter $\theta$ in the boolean oracle case, say $\theta_\mathrm{bool}$. Let $P^\mathrm{good}_0$ be the probability for a measurement on the initial state $\Ket{\psi_0}$ to yield a winning state. From \eqref{eq:costheta},
\begin{align}
 \cos(\theta_\mathrm{bool}) &= \left[-1 \times P^\mathrm{good}_0\right] + \left[1 \times \left(1-P^\mathrm{good}_0\right)\right] = 1 - 2 P^\mathrm{good}_0\,,\label{eq:theta_bool_1}\\
 \Rightarrow \sin^2\left(\theta_\mathrm{bool}/2\right) &= P^\mathrm{good}_0\,.\label{eq:theta_bool_2}
\end{align}
Thus, in the boolean oracle case
\begin{itemize}
 \item $\sin^2(\theta/2)$ reduces to the initial probability of ``success'' (measuring a winning state), which is captured by $\sin^2(\theta_a)$ in Ref.~\cite{QAAAE}, and
 \item The parameter $\theta$ used in this paper reduces to the parameter $2\theta_a$ used in Ref.~\cite{QAAAE}, and $\lfloor\pi/(2\theta)\rfloor$ reduces to $\lfloor\pi/(4\theta_a)\rfloor$.
\end{itemize}
In this way, the results of \sref{sec:anal} in general, and \sref{subsec:optimalK} in particular, can be seen as generalizations of the corresponding results in Ref.~\cite{QAAAE}.

\subsection{Alternative Formulation of the Algorithm} \label{subsec:alternative_form}
In the formulation of the non-boolean amplitude amplification algorithm in \sref{sec:nbaaa}, an ancilla qubit (first register) was included solely for the purpose of making the quantity $\Braket{\Psi_0|\Uphi|\Psi_0} = \Braket{\Psi_0|\Alpha}$ real-valued. If, in a particular use case, it is guaranteed that $\Braket{\psi_0|U_\varphi|\psi_0}$ will be real-valued (or have a negligible imaginary part\footnote{This could be achieved, e.g, by replacing the function $\varphi(x)$ with $\varphi'(x) = r(x)\varphi(x)$, where $r:\{0,1,\dots N-1\} \rightarrow \{-1, +1\}$ is a random function independent of $\varphi(x)$, with mean $0$ (for $x$ sampled by measuring $\Ket{\psi_0}$).}) even without introducing the ancilla, then the algorithm described in \sref{sec:nbaaa} can be used without the ancilla: alternate between applying $S_{\psi_0}\,U_\varphi$ during the odd iterations and $S_{\psi_0}\,U^\dagger_\varphi$ during the even iterations. In other words, the properties and structure of two-register system were not exploited in the algorithm description, beyond making $\Braket{\Psi_0|\Uphi|\Psi_0}$ real-valued.

However, from \eqref{eq:alpha_expand} and \eqref{eq:beta_expand}, it can be seen that the states $\Ket{\Alpha} = \Uphi\Ket{\Psi_0}$ and $\Ket{\Beta} = \Uphi^\dagger\Ket{\Psi_0}$ are related by
\begin{equation}
 \Ket{\Beta} = [X\otimes I]\Ket{\Alpha}\,,\qquad\qquad \Ket{\Alpha} = [X\otimes I]\Ket{\Beta}\,, \label{eq:alphatobeta}
\end{equation}
where $X$ is the bit-flip or the Pauli-X operator. This can be exploited to avoid having two separate cases---$k$ being odd and even---in the final expression for $\Ket{\Psi_k}$ in \eqref{eq:Psik_full}. The expression for both cases can be made the same by acting the Pauli-X gate on the ancilla, once at the end, if the total number of iterations is even.

More interestingly, the relationship between $\Ket{\Alpha}$ and $\Ket{\Beta}$ can be used to avoid having two different operations in the first place, for the odd and even iterations. This leads to the following alternative formulation of the non-boolean amplitude amplification algorithm: During each iteration, odd or even, act the same operator $\Uiter$ defined by
\begin{equation}
 \Uiter \equiv \Udif\,\Uphi\,[X\otimes I]\,. \label{eq:uiter}
\end{equation}
This alternative formulation is depicted as a circuit in \fref{alg:algorithm_alt} and as a pseudocode in \alref{alg:algorithm_alt}.
\begin{figure}[t]
 \centering
 \begin{tikzpicture}
  \node[scale=1.0] {
   \begin{quantikz}
    \lstick{$\displaystyle\genfrac{}{}{0pt}{}{\text{register 1}}{\text{(ancilla)}}$ : $\Ket{+}$} & \qw & \qw & \gate{X} \gategroup[wires=2,steps=3,style={dashed,rounded corners,fill=blue!10,inner xsep=10pt,inner ysep=5pt,outer ysep=3pt},background]{$\Uiter$ (single iteration)} & \gate[wires=2]{\Uphi} & \gate[wires=2]{\Udif} & \qw & ~\ldots\qw~ & \meter{0/1} & \qw \rstick{~(ignored)} \\[1em]
    \lstick{register 2: $\Ket{\psi_0}$} & \qwmult & \qwmult & \qwmult & \qwmult & \qwmult & \qwmult & ~\ldots\qwmult~ & \qwmult & \qwmult \rstick{~Final state}
  \end{quantikz}
  };
 \end{tikzpicture}
 \caption{Quantum circuit for the alternative formulation, in \sref{subsec:alternative_form}, of the non-boolean amplitude amplification algorithm.}
 \label{fig:algorithm_alt}
\end{figure}
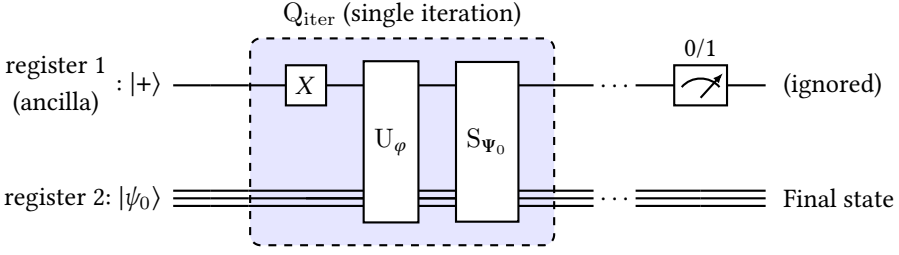
\begin{algorithm}[t]
 \begin{algorithmic}[1]
  \STATE \textbf{initialize} $\Ket{\Psi} := \Ket{\Psi_0}$
  \FOR{$k := 1$ \textbf{to} $K$}
   \STATE \textbf{update} $\Ket{\Psi} := \Udif\,\Uphi\,[X\otimes I] \Ket{\Psi}$
  \ENDFOR
  \STATE Measure the ancilla in the 0/1 basis.
 \end{algorithmic}
 \caption{Alternative formulation, in \sref{subsec:alternative_form}, of the non-boolean amplitude amplification algorithm.} \label{alg:algorithm_alt}
\end{algorithm}

From \eqref{eq:iter_iden_start}, \eqref{eq:iter_iden_2}, and \eqref{eq:alphatobeta}, the action of $\Uiter$ on $\Ket{\Psi_0}$ and $\Ket{\Alpha}$ can be derived as follows:
\begin{alignat}{3}
 \Uiter &\Ket{\Psi_0} &&= \Udif\,\Uphi \Ket{\Psi_0} &&= 2\cos(\theta) \Ket{\Psi_0} - \Ket{\Alpha}\,, \label{eq:uiter_psi0}\\
 \Uiter &\Ket{\Alpha} &&= \Udif\,\Uphi \Ket{\Beta} &&= \Ket{\Psi_0}\,. \label{eq:uiter_alpha}
\end{alignat}
Let $\bigket{\Psi^\mathrm{alt}_k}$ be the state of the two-register system after $k$ iterations under this alternative formulation (before any measurement of the ancilla).
\begin{equation}
 \bigket{\Psi^\mathrm{alt}_{k}} \equiv \Uiter^k \Ket{\Psi_0}\,.
\end{equation}
Using similar manipulations as in \aref{appendix:state_derivation}, $\bigket{\Psi^\mathrm{alt}_k}$ can be expressed, for all $k\geq0$, as
\begin{equation}
 \bigket{\Psi^\mathrm{alt}_k} = \frac{1}{\sin(\theta)} \Big[\sin\!\big((k+1)\theta\big) \Ket{\Psi_0} -\sin(k\theta) \Ket{\Alpha}\Big]\,. \label{eq:Psik_prime}
\end{equation}
Note that this expression for $\bigket{\Psi^\mathrm{alt}_k}$ is almost identical to the expression for $\Ket{\Psi_k}$ in \eqref{eq:Psik_full}, but without two separate cases for the odd and even values of $k$. Much of the analysis of the original formulation of the algorithm in \sref{sec:anal} holds for the alternative formulation as well, including the expressions for the state probabilities $p_K(x)$, mean $\mu_K$, raw moments $\mu^{(n)}_K$, and the cumulative distribution function $F^{\cos}_K$. 

In addition to simplifying the amplification algorithm (by using the same operation for every iteration), the $\Uiter$ operator used in this subsection allows for a clearer presentation of the quantum mean estimation algorithm, which will be introduced next.

\section{Quantum Mean Estimation Algorithm} \label{sec:mean}
The goal of the quantum mean estimation algorithm is to estimate the expected value of $e^{i\varphi(x)}$ for $x$ sampled by measuring a given superposition state $\Ket{\psi_0}$. Let $E_{\psi_0}\!\big[e^{i\varphi}\big]$ denote this expected value.
\begin{equation}
 E_{\psi_0}\!\big[e^{i\varphi}\big]\equiv\Braket{\psi_0|U_\varphi|\psi_0} = \sum_{x=0}^{N-1}\big|a_0(x)\big|^2\, e^{i\varphi(x)}\,.
\end{equation}
This can be written as
\begin{equation}
  E_{\psi_0}\!\big[e^{i\varphi}\big] = \RE\Big[E_{\psi_0}\!\big[e^{i\varphi}\big]\Big] + i~ \IM\Big[E_{\psi_0}\!\big[e^{i\varphi}\big]\Big]\,,
\end{equation}
where the real and imaginary parts are given by
\begin{align}
 \RE\Big[E_{\psi_0}\!\big[e^{i\varphi}\big]\Big] &= \sum_{x=0}^{N-1} \big|a_0(x)\big|^2\cos\!\big(\varphi(x)\big)\,,\\
 \IM\Big[E_{\psi_0}\!\big[e^{i\varphi}\big]\Big] &= \sum_{x=0}^{N-1} \big|a_0(x)\big|^2\sin\!\big(\varphi(x)\big)\,.
\end{align}
The mean estimation can therefore be performed in two parts---one for estimating the mean of $\cos(\varphi)$, and the other for estimating the mean of $\sin(\varphi)$. Note that the expectation of $\cos(\varphi)$ under the state $\Ket{\psi_0}$ is precisely $\cos(\theta)$ defined in \eqref{eq:costheta}.

\subsection{Estimating the Mean of \texorpdfstring{$\cos(\varphi)$}{cos(phi)}} \label{subsec:mean_eigen_re}

The connection shown in \sref{subsec:boolean_case} between the parameter $\theta$ and the parameter $\theta_a$ used in Ref.~\cite{QAAAE} serves as the intuition behind the quantum mean estimation algorithm of this paper. In the amplitude estimation algorithm of Ref.~\cite{QAAAE}, the parameter $\theta_a$ is estimated\footnote{The estimation of $\theta_a$ is only (needed to be) performed up to a two-fold ambiguity of $\{\theta_a, \pi-\theta_a\}$.} using QPE. The estimate for $\theta_a$ is then turned into an estimate for the initial winning probability. Likewise, here the basic idea is to estimate\footnote{The estimation of $\theta$ is only (needed to be) performed up to a two-fold ambiguity of $\{\theta, 2\pi-\theta\}$.} the parameter $\theta$ defined in \eqref{eq:costheta} using QPE. The estimate for $\theta$ can then be translated into an estimate for the initial expected value of $\cos\!\big(\varphi(x)\big)$, namely $\cos(\theta)$. The rest of this subsection will actualize this intuition into a working algorithm.



The key observation\footnote{The forms of the eigenstates $\Ket{\Eta_+}$ and $\Ket{\Eta_-}$ in \eqref{eq:eta} can be guessed from the form of the matrix $S_\theta$ in \aref{appendix:state_derivation}. They can also be guessed from \eqref{eq:Psik_prime}, by rewriting the $\sin$ functions in terms of complex exponential functions.} is that $\Ket{\Psi_0}$ can be written as
\begin{equation}
 \Ket{\Psi_0} = \frac{\Ket{\Eta_+} - \Ket{\Eta_-}}{\sqrt{2}}\,, \label{eq:Psi0_eta}
\end{equation}
where $\Ket{\Eta_+}$ and $\Ket{\Eta_-}$ are given by
\begin{equation}
 \Ket{\Eta_\pm} = \frac{e^{\pm i\theta}\Ket{\Psi_0} - \Ket{\Alpha}}{i\,\sqrt{2}\,\sin(\theta)}\,. \label{eq:eta}
\end{equation}
The expressions for $\Ket{\Eta_+}$ and $\Ket{\Eta_-}$ in \eqref{eq:eta} can be used to verify \eqref{eq:Psi0_eta}. Crucially, $\Ket{\Eta_+}$ and $\Ket{\Eta_-}$ are unit normalized eigenstates of the unitary operator $\Uiter$, with eigenvalues $e^{i\theta}$ and $e^{-i\theta}$, respectively.
\begin{align}
 \Uiter\,\Ket{\Eta_\pm} &= e^{\pm i\theta}\,\Ket{\Eta_\pm}\,, \label{eq:eta_action}\\
 \Braket{\Eta_\pm|\Eta_\pm} &= 1\,. \label{eq:eta_norm}
\end{align}
The properties of $\Ket{\Eta_+}$ and $\Ket{\Eta_-}$ in \eqref{eq:eta_action} and \eqref{eq:eta_norm} can be verified using \eqref{eq:uiter_psi0}, \eqref{eq:uiter_alpha}, and \eqref{eq:eta}, as shown in \aref{appendix:uiter_eigen}. The observations in \eqref{eq:Psi0_eta} and \eqref{eq:eta_action} lead to the following algorithm for estimating $\cos(\theta)$:
\begin{enumerate}
 \item Perform the QPE algorithm with
 \begin{itemize}
  \item the two-register operator $\Uiter$ serving the role of the unitary operator under consideration, and
  \item the superposition state $\Ket{\Psi_0}$ in place of the eigenstate required by the QPE algorithm as input.
 \end{itemize}
 Let the output of this step, appropriately scaled to be an estimate of the phase angle in the range $[0,2\pi)$, be $\hat{\omega}$.
 \item Return $\cos\!\big(\hat{\omega}\big)$ as the estimate for $\cos(\theta)$, i.e., the real part of $E_{\psi_0}\!\big[e^{i\varphi}\big]$.\footnote{If the circuit implementation of $\Uiter$ is wrong by an overall (state independent) phase $\phi_\mathrm{err}$, then the estimate for $\cos(\theta)$ is $\cos(\hat{\omega}-\phi_\mathrm{err})$. This is important, for example, if the operation $\big[2\Ket{0,0}\Bra{0,0}-\I\big]$ is only implemented up to a factor of $-1$, i.e., with $\phi_\mathrm{err} = \pi$. Note that the final state probabilities under the non-boolean amplitude amplification algorithm are unaffected by such an overall phase error.}
\end{enumerate}
\textbf{Proof of correctness of the algorithm:} $\Ket{\Psi_0}$ is a superposition of the eigenstates $\Ket{\Eta_+}$ and $\Ket{\Eta_-}$ of the unitary operator $\Uiter$. This implies that $\hat{\omega}$ will either be an estimate for the phase angle of $\Ket{\Eta_+}$, namely $\theta$, or an estimate for the phase angle of $\Ket{\Eta_-}$, namely $2\pi-\theta$.\footnote{If $\theta=0$, the phase angle being estimated is $0$ for both $\Ket{\Eta_+}$ and $\Ket{\Eta_-}$.}\textsuperscript{,}%
\footnote{For $\theta\neq0$, $\hat{\omega}$ will be an estimate for either $\theta$ or $2\pi-\theta$ with \emph{equal probability}, but this detail is not important.}\textsuperscript{,}%
\footnote{Since $\theta$ lies in $[0,\pi]$ and $2\pi-\theta$ lies in $[\pi, 2\pi]$, the output $\hat{\omega}$ can be converted into an estimate for $\theta$ alone. But this is not necessary.} Since, $\cos(2\pi-\theta) = \cos(\theta)$, it follows that $\cos\!\big(\hat{\omega}\big)$ is an estimate for $\cos(\theta)$.

\subsection{Estimating the Mean of \texorpdfstring{$e^{i\varphi}$}{exp(i phi)}} \label{subsec:mean_eigen}
The algorithm for estimating the expected value of $\cos(\varphi)$ in the previous subsection can be re-purposed to estimate the expected value of $\sin(\varphi)$ by using the fact that
\begin{equation}
 \sin(\varphi) = \cos(\varphi-\pi/2)\,.
\end{equation}
In other words, the imaginary part of $E_{\psi_0}\!\big[e^{i\varphi}\big]$ is the real part of $E_{\psi_0}\!\big[e^{i(\varphi-\pi/2)}\big]$. By using the oracle $\U_{\varphi-\pi/2}$ (for the function $\varphi-\pi/2$), instead of $\Uphi$, in the mean estimation algorithm of \sref{subsec:mean_eigen_re}, the imaginary part of $E_{\psi_0}\!\big[e^{i\varphi}\big]$ can also be estimated. This completes the estimation of $E_{\psi_0}\!\big[e^{i\varphi}\big]$.

\begin{figure}[t]
 \centering
 \begin{tikzpicture}
  \node[scale=1.0] {
   \begin{quantikz}
    \lstick{$\displaystyle\genfrac{}{}{0pt}{}{\text{register 1}}{\text{(ancilla)}}$} & \qw & \gate[wires=2]{\Uphi}\gategroup[wires=2,steps=5,style={dashed,rounded corners,inner xsep=5pt,inner ysep=5pt,outer ysep=5pt},background]{Circuit for \,$\U_{\varphi-\pi/2}$} & \gate{R_{\pi/2}} & \gate{X} & \gate{R_{-\pi/2}} & \gate{X} & \qw & \qw \\[1em]
    \lstick{register 2} & \qwmult & \qwmult & \qwmult & \qwmult & \qwmult & \qwmult & \qwmult & \qwmult
   \end{quantikz}
  };
 \end{tikzpicture}
 \caption{Quantum circuit for an implementation of $\U_{\varphi-\pi/2}$ using $\Uphi$.}
 \label{fig:Uphi_minuspiby2}
\end{figure}
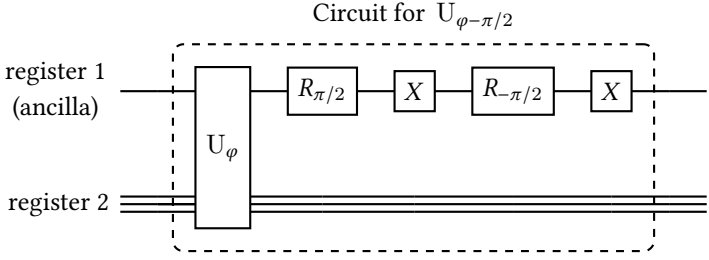

For concreteness, $\U_{\varphi-\pi/2}$ can be explicitly written as
\begin{align}
 \U_{\varphi-\pi/2} = e^{-i\pi/2}\Ket{0}\Bra{0} \otimes U_{\varphi} + e^{i\pi/2}\Ket{1}\Bra{1} \otimes U^\dagger_{\varphi}\,.
\end{align}
An implementation of $\U_{\varphi-\pi/2}$ using the oracle $\Uphi$, the bit-flip operator $X$, and the phase-shift operator $R_\phi$ is shown in \fref{fig:Uphi_minuspiby2}.

Note that the algorithm does not, in any way, use the knowledge that $\{\Ket{0},\dots,\Ket{N-1}\}$ is an eigenbasis of $U_\varphi$. So, this algorithm can be used to estimate $\Braket{\psi_0|U|\psi_0}$ for any unitary operator $U$.

\subsection{Quantum Speedup}
The speedup offered by the quantum mean estimation algorithm over classical methods will be discussed here, in the context of estimating the mean of $\cos(\varphi)$ alone. The discussion can be extended in a straightforward way to the estimation of $E_{\psi_0}\!\big[e^{i\varphi}\big]$.
\subsubsection{Classical Approaches to Estimating the Mean}

For an arbitrary function $\varphi$ and a known sampling distribution $p_0(x)$ for the inputs $x$, one classical approach to finding the mean of $\cos\!\big(\varphi(x)\big)$ is to sequentially query the value of $\varphi(x)$ for all the inputs, and use the query results to compute the mean. Let the permutation $(x_0, x_1, \dots, x_{N-1})$ of the inputs $(0,1,\dots, N-1)$ be the order in which the inputs are queried. The range of allowed values of $\cos(\theta)$, based only on results for the first $q$ inputs, is given by
\begin{equation}
 \left[\sum_{j=0}^{q-1} p_0(x)\,\cos\!\big(\varphi(x_j)\big) - \sum_{j=q}^{N-1} p_0(x)\right] \leq \cos(\theta) \leq \left[\sum_{j=0}^{q-1} p_0(x)\,\cos\!\big(\varphi(x_j)\big) + \sum_{j=q}^{N-1} p_0(x)\right]\,.
\end{equation}
These bounds are derived by setting the values of $\cos(\varphi)$ for all the unqueried inputs to their highest and lowest possible values, namely $+1$ and $-1$. The range of allowed values of $\cos(\theta)$ shrinks as more and more inputs are queried. In particular, if $p_0(x)$ is equal for all the inputs $x$, the width of the allowed range (based on $q$ queries) is given by $2(N-q)/N$. This strategy will take $\mathcal{O}(N)$ queries before the width of the allowed range reduces to even, say, 1. Thus, this strategy will not be feasible for large values of $N$.

A better classical approach is to probabilistically estimate the expected value as follows:
\begin{enumerate}
 \item Independently sample $q$ random inputs $(x_1,\dots x_q)$ as per the distribution $p_0$.
 \item Return the sample mean of $\cos(\varphi)$ over the random inputs as an estimate for $\cos(\theta)$.
\end{enumerate}
Under this approach, the standard deviation of the estimate scales as $\sim\!\sigma_0/\sqrt{q}$, where $\sigma_0$ is the standard deviation of $\cos(\varphi)$ under the distribution $p_0$.
\subsubsection{Precision Vs Number of Queries for the Quantum Algorithm}

Note that one call to the operator $\Uiter$ corresponds to $\mathcal{O}(1)$ calls to $A_0$ and the oracles $U_\varphi$ and $U_\varphi^\dagger$. Let $q$ be the number of times the (controlled) $\Uiter$ operation is performed during the QPE subroutine. As $q$ increases, the uncertainty on the estimate for the phase-angle $\theta$ (up to a two-fold ambiguity) falls at the rate of $\mathcal{O}(1/q)$ \cite{doi:10.1098/rspa.1998.0164}. Consequently, the uncertainty on $\cos(\theta)$ also falls at the rate of $\mathcal{O}(1/q)$. This represents a \textbf{quadratic speedup} over the classical, probabilistic approach, under which the error falls as $\mathcal{O}(1/\sqrt{q})$. Note that the variance of the estimate for $\cos(\theta)$ is independent of a) the size of input space $N$, and b) the variance $\sigma^2_0$ of $\cos\!\big(\varphi(x)\big)$ under the distribution $p_0(x)$. It only depends on the true value of $\cos(\theta)$ and the number of queries $q$ performed during the QPE subroutine.

\section{Demonstrating the Algorithms Using a Toy Example} \label{sec:toy}
In this section, the non-boolean amplitude amplification algorithm and the mean estimation algorithm will both be demonstrated using a toy example. Let the input to the oracle $U_\varphi$, i.e., the second register, contain 8 qubits. This leads to $2^8 = 256$ basis input states, namely $\Ket{0}$, \dots, $\Ket{255}$. Let the toy function $\varphi(x)$ be
\begin{equation}
 \varphi(x) = \frac{x}{255}\,\frac{\pi}{4}\,,\qquad\qquad \text{for } x=0,1,\dots,255\,.
\end{equation}
The largest phase-shift applied by the corresponding oracle $U_\varphi$ on any basis state is $\pi/4$, for the state $\Ket{255}$. Since, $\cos\!\big(\varphi(x)\big)$ is monotonically decreasing in $x$, the goal of the amplitude amplification algorithm is to amplify the probabilities of higher values of $x$.

Let the initial state, from which the amplification is performed, be the uniform superposition state $\Ket{s}$.
\begin{equation}
 \Ket{\psi_0} = \Ket{s} = \frac{1}{\sqrt{256}}~\sum_{x=0}^{255}\Ket{x}\,.
\end{equation}
Such simple forms for the oracle function and the initial state allow for a good demonstration of the algorithms.

\begin{figure}
 \centering
 \includegraphics[width=.75\textwidth]{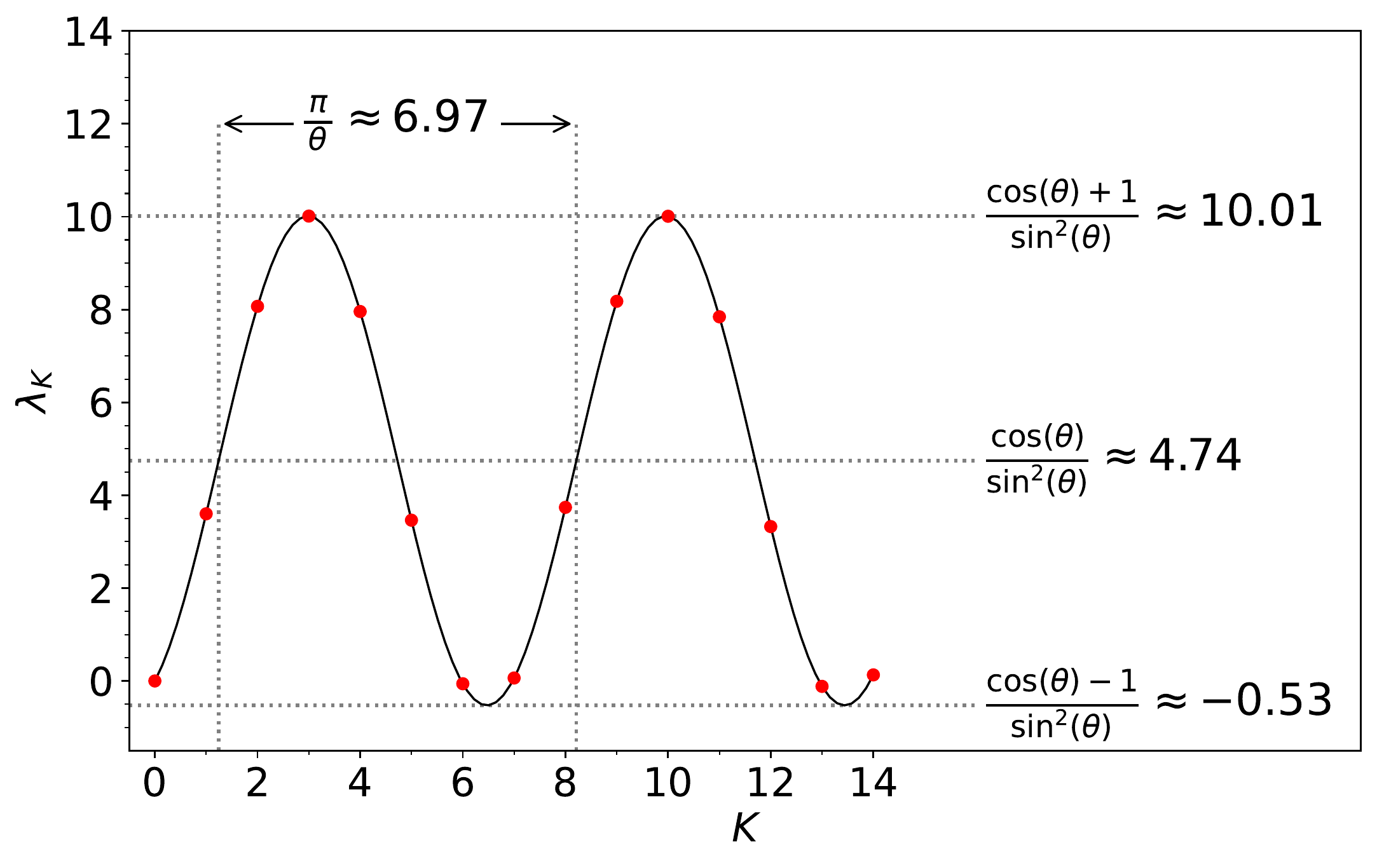}
 \caption{Plot showing $\lambda_K$ for $K=0,1,\dots,14$, for the toy example considered in \sref{sec:toy}. The red dots correspond to the different integer values of $K$. The black solid curve depicts the sinusoidal dependence of $\lambda_K$ on $K$. The dotted lines indicate that $\lambda_K$ oscillates around $\cos(\theta)/\sin^2(\theta)$ with an amplitude of $1/\sin^2(\theta)$ and a period of $\pi/\theta$ (in $K$ values).}
 \label{fig:lambda}
\end{figure}

For this toy example, from \eqref{eq:costheta}, $\cos(\theta)$ and $\theta$ are given by
\begin{align}
 \cos(\theta) = \frac{1}{256} \sum_{x=0}^{255} \cos\left(\frac{x}{255}\,\frac{\pi}{4}\right) &\approx 0.9001\,,\\
 \theta &\approx 0.4507\,.
\end{align}
\Fref{fig:lambda} shows the value of $\lambda_K$, from \eqref{eq:lambda}, for the first few values of $K$.
The heuristic choice for the total number of iterations $\tilde{K} = \lfloor\pi/(2\theta)\rfloor$ is 3 for this example, as can also be seen from \fref{fig:lambda}.

\subsection{Amplitude Amplification}
\begin{figure}
 \centering
 \includegraphics[width=.75\textwidth]{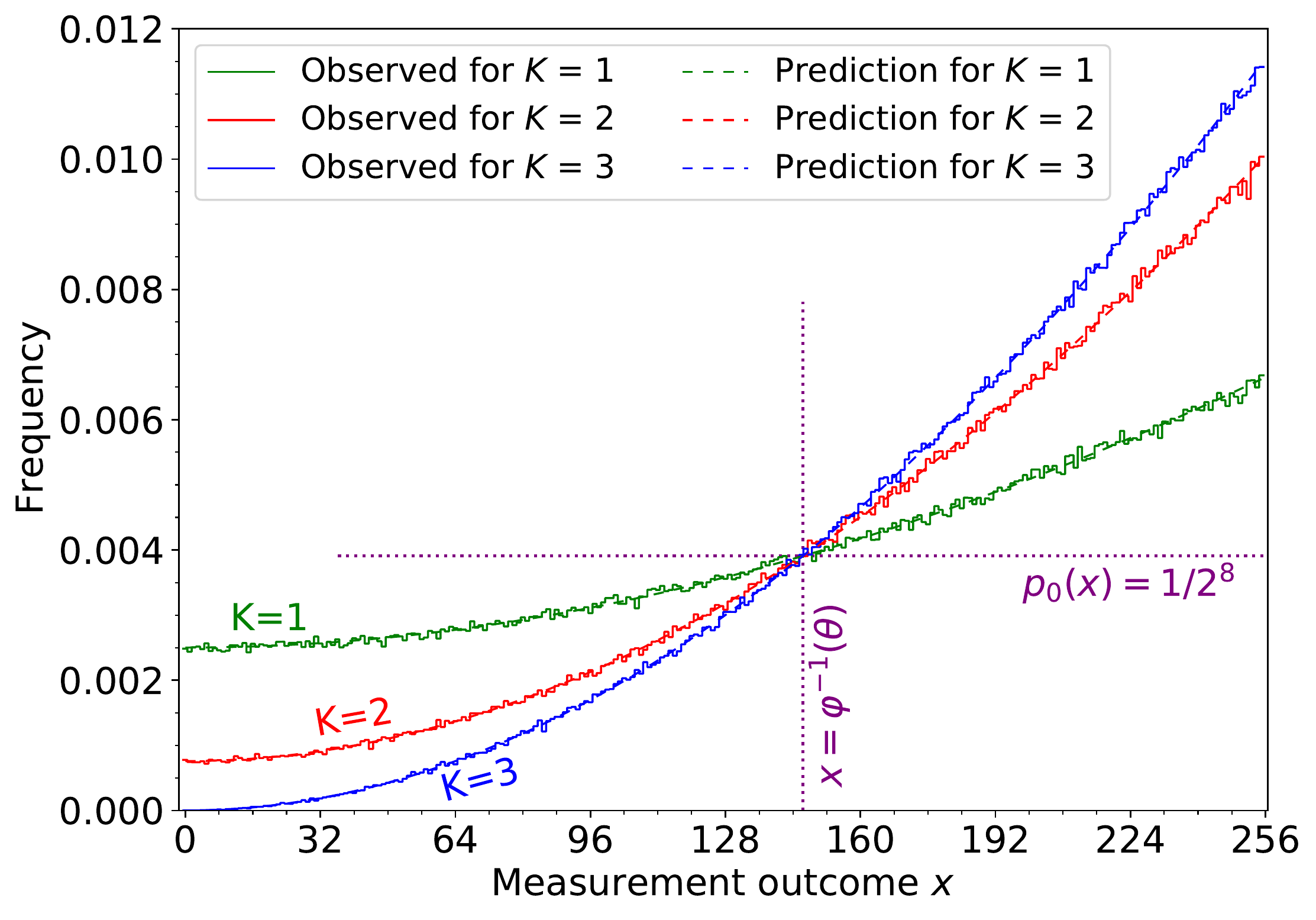}
 \caption{The solid histograms show the observed measurement frequencies of the different values of $x\in\{0,1\dots,255\}$ after performing the non-boolean amplitude amplification algorithm. The green, red, and blue solid histograms correspond to the total number of iterations $K$ being 1, 2, and 3, respectively. In each case, the observed frequencies are based on simulating the circuit for the algorithm $10^6$ times, i.e., $10^6$ shots. The dashed curves, almost coincident with their corresponding solid histograms, show the predictions $p_K(x)$ (for the measurement frequencies) computed using \eqref{eq:Pk_short}. While $p_K(x)$ is technically defined only for the integer values of $x$, here the dashed curves are interpolated for non-integer values of $x$ using \eqref{eq:Pk_short}.}
 \label{fig:histograms}
\end{figure}

For this toy example, the quantum circuit for the non-boolean amplitude amplification algorithm was implemented in Qiskit \cite{Qiskit} for three different values of the total number of iterations $K$, namely $K=1, 2, \text{ and }3$. In each case, the resulting circuit was simulated (and measured) $10^6$ times using Qiskit's QASM simulator. The estimated measurement frequencies for the observations $x=0,1,\dots,255$ are shown in \fref{fig:histograms} as solid, unfilled, histograms---the colors green, red, and blue correspond to $K=1$, $2$, and $3$, respectively. The expected measurement frequencies, namely $p_K(x)$ from \eqref{eq:Pk_short}, are also shown in \fref{fig:histograms} as dashed curves, and are in good agreement with the corresponding histograms.

As can be seen from \fref{fig:histograms}, in each case, the algorithm preferentially amplifies lower values of $\cos\!\big(\varphi(x)\big)$ or, equivalently, higher values of $x$. This is expected from the fact that $\lambda_K>0$ for all three values of $K$. Furthermore, as $K$ increases from 0 to $\tilde{K}=3$, the preferential amplification grows stronger. Note that the probabilities of the $x$-s for which $\cos\!\big(\varphi(x)\big)\approx\cos(\theta)$ are left approximately unchanged by the algorithm, as indicated by the purple-dotted crosshair in \fref{fig:histograms}.

\subsection{Mean Estimation}
Only the estimation of $\cos(\theta)$, i.e., the real part of $E\!\left[e^{i\varphi}\right]$ is demonstrated here. The imaginary part can also be estimated using the same technique, as described in \sref{subsec:mean_eigen}.

Let $M$ be the number of qubits used in the QPE subroutine of the mean estimation algorithm, to contain the phase information. This corresponds to performing the (controlled) $\Uiter$ operation $2^M-1$ times during the QPE subroutine. Note that the estimated phase $\hat{\omega}$ can only take the following discrete values \cite{doi:10.1098/rspa.1998.0164}:
\begin{equation}
 \hat{\omega} \in \left\{~~\frac{2\pi j}{2^{M}}\quad\Big|\quad j\in\left\{0,\dots,2^M-1\right\}~~\right\}\,.
\end{equation}
In this way, the value of $M$ controls the precision of the estimated phase and, by extension, the precision of the estimate for $\cos(\theta)$---the higher the value of $M$, the higher the precision.

Two different quantum circuits were implemented, again using Qiskit, for the mean estimation algorithm; one with $M=4$ and the other with $M=8$. Each circuit was simulated (and measured) using Qiskit's QASM simulator $10^6$ times, to get a sample of $\hat{\omega}$ values, all in the range $[0, 2\pi)$.

\begin{figure}
 \centering
 \includegraphics[width=.47\textwidth]{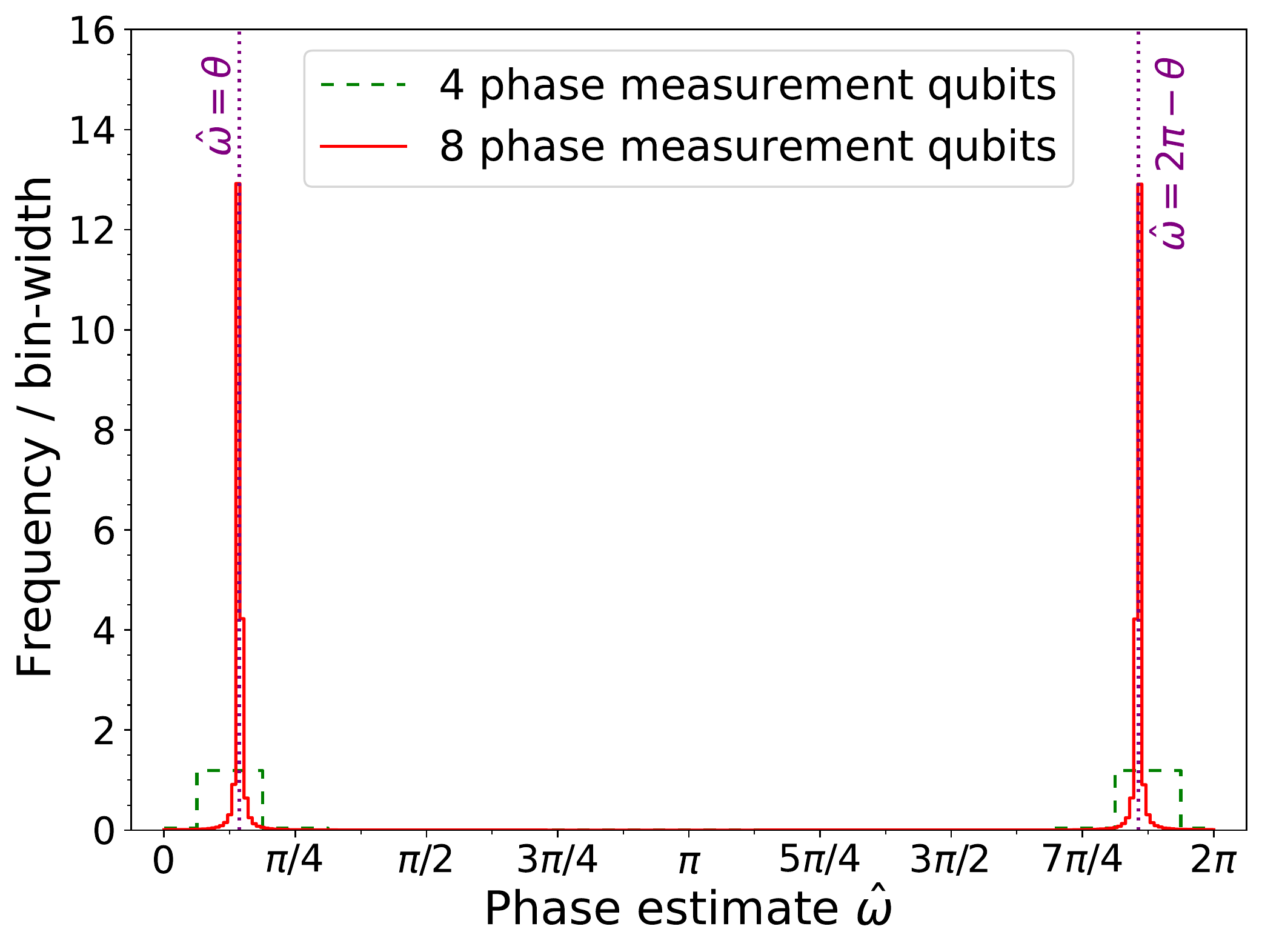}\quad
 \includegraphics[width=.47\textwidth]{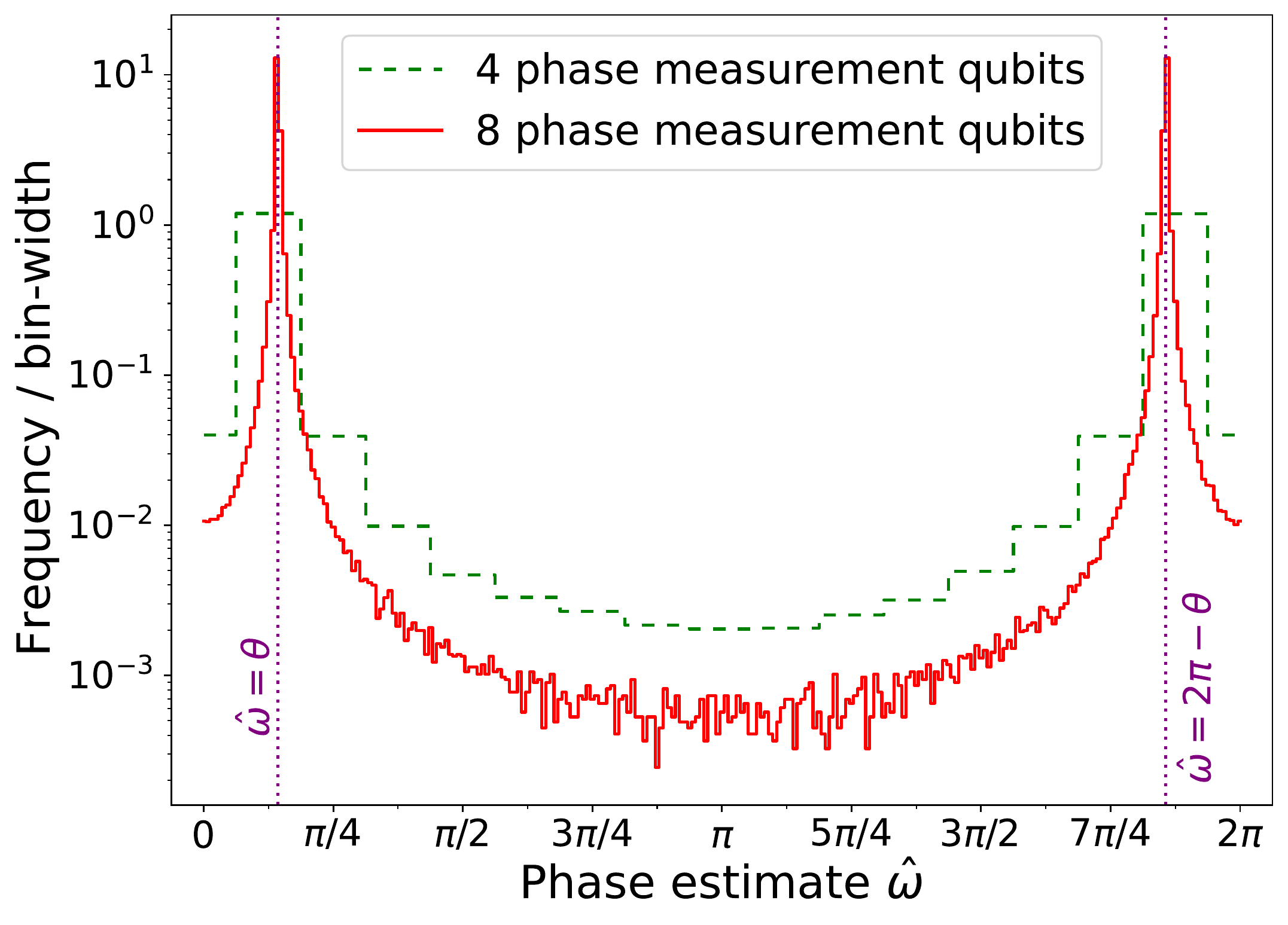}
 \caption{Histograms showing the observed frequency/bin-width for the different phase estimation outcomes $\hat{\omega}$, for the toy example considered in \sref{sec:toy}. The left and right panels show the frequency/bin-width ($y$-axis) on linear and logarithmic scales, respectively. In both panels, the green-dashed and red-solid histograms correspond to using $4$ and $8$ qubits to measure the phase, respectively. The purple-dotted vertical lines on both panels correspond to $\hat{\omega}=\theta$ and $\hat{\omega}=2\pi-\theta$. In each histogram, the last bin (ending at $2\pi$) is simply a continuation of the first bin (starting at $0$), and corresponds to $\hat{\omega}=0$.}
 \label{fig:phase_histograms}
\end{figure}

The observed frequencies (scaled by 1/bin-width) of the different values of $\hat{\omega}$ are shown as histograms on a linear scale in the left panel of \fref{fig:phase_histograms}, and on a logarithmic scale in the right panel. Here the bin-width of the histograms is given by $2\pi/2^M$, which is the difference between neighboring allowed values of $\hat{\omega}$.
The green-dashed and red-solid histograms in \fref{fig:phase_histograms} correspond to the circuits with $4$ and $8$ phase measurement qubits, respectively. The exact values of $\theta$ and $2\pi-\theta$ for this toy example are indicated with vertical purple-dotted lines. In both cases ($M=4$ and $M=8$), the observed frequencies peak near the exact values of $\theta$ and $2\pi-\theta$, demonstrating that $\hat{\omega}$ is a good estimate for them, up to a two-fold ambiguity. Furthermore, as expected, using more qubits for estimating the phase leads to a more precise estimate.

\begin{figure}
 \centering
 \includegraphics[width=.47\textwidth]{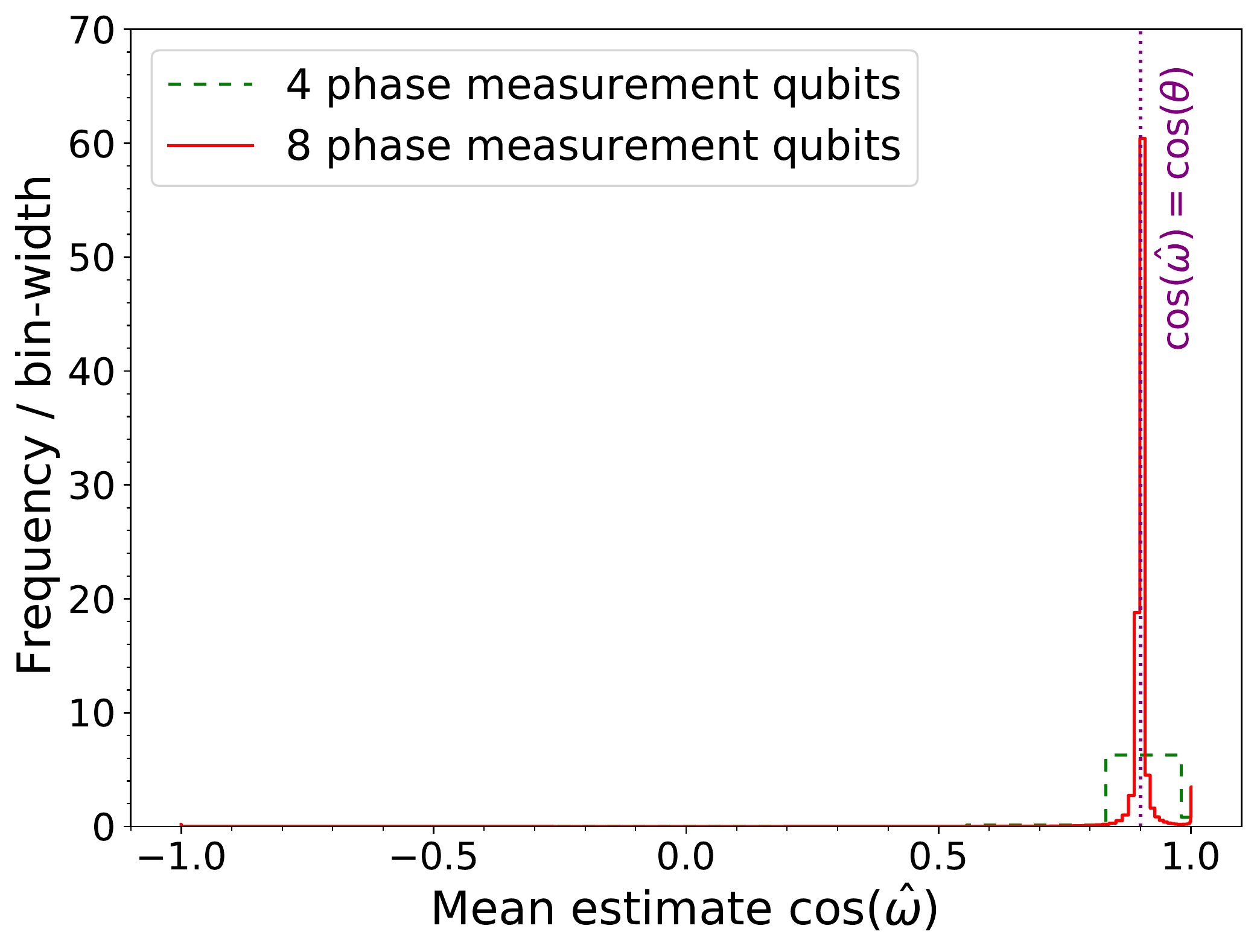}\quad
 \includegraphics[width=.47\textwidth]{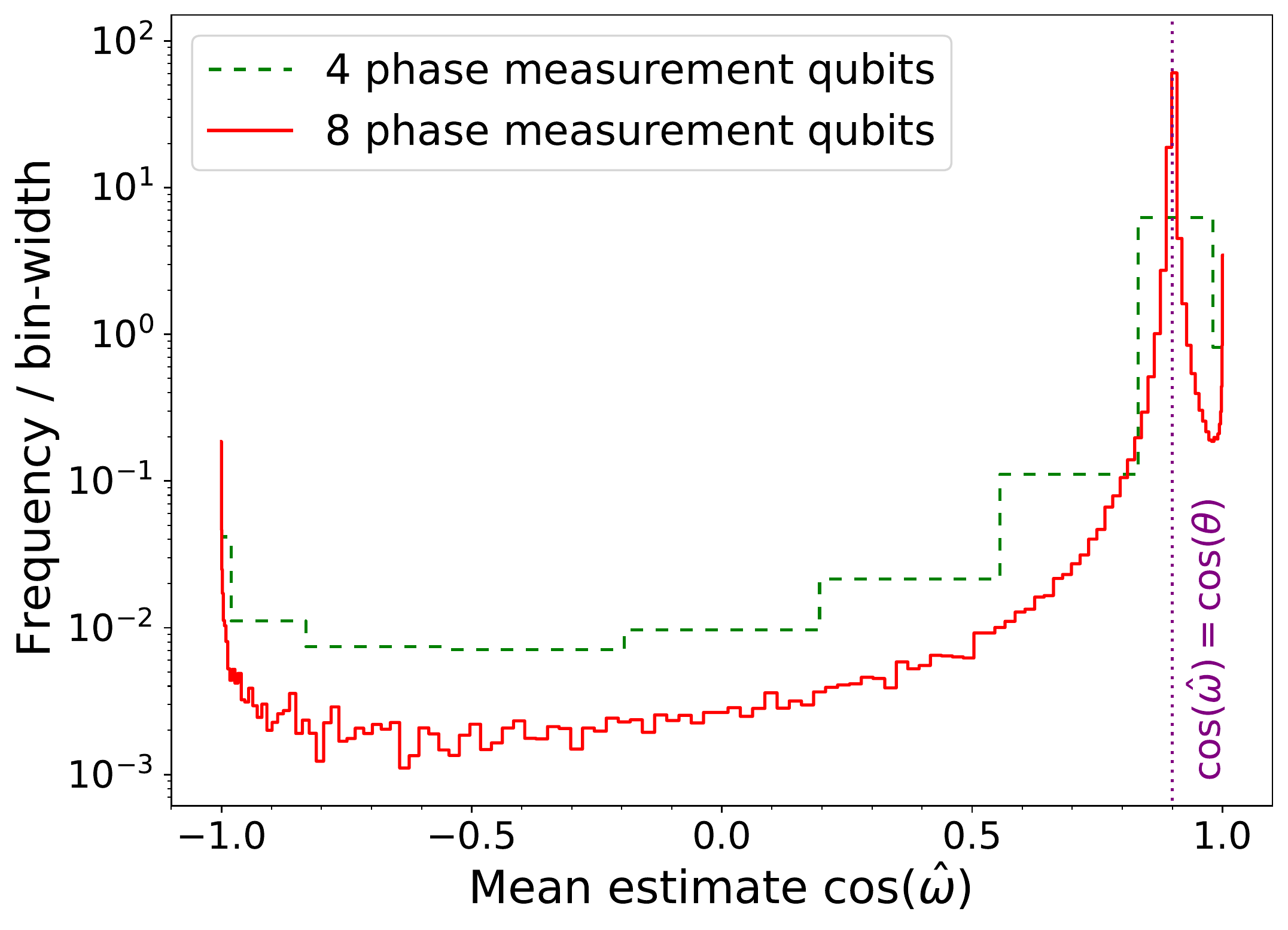}
 \caption{Histograms showing the observed frequency/bin-width for the different values of $\cos(\hat{\omega})$, for the toy example considered in \sref{sec:toy}. The left and right panels show the frequency/bin-width ($y$-axis) on linear and logarithmic scales, respectively. In both panels, the green-dashed and red-solid histograms correspond to using 4 and 8 qubits respectively. The purple-dotted vertical lines correspond to $\cos(\hat{\omega}) = \cos(\theta')$. The bins of the $\cos(\hat{\omega})$ histograms are induced by the equal-width bins of the corresponding $\hat{\omega}$ histograms in \fref{fig:phase_histograms}.}
 \label{fig:cosphase_histograms}
\end{figure}

\Fref{fig:cosphase_histograms} shows the observed frequencies (scaled by 1/bin-width) of the different values of $\cos(\hat{\omega})$, which is the estimate for the mean $\cos(\theta)$. As with \fref{fig:phase_histograms}, a) the green-dashed and red-solid histograms correspond to 4 and 8 phase measurement qubits, respectively, and b) the left and right panels show the histograms on linear and logarithmic scales, respectively. In each panel, the exact value of $\cos(\theta)$ is also indicated as a vertical purple-dotted line. As can be seen from \fref{fig:cosphase_histograms}, the observed frequencies peak\footnote{The upward trends near the left ($-1$) and right ($+1$) edges of the plots in \fref{fig:cosphase_histograms} are artifacts caused by the Jacobian determinant for the map from $\hat{\omega}$ to $\cos(\hat{\omega}$).} near the exact value of $\cos(\theta)$, indicating that $\cos(\hat{\omega})$ is a good estimate for the same.

\section{Ancilla-Free Versions of the Algorithms} \label{sec:no_ancilla}
Both algorithms introduced this in this paper so far, namely
\begin{itemize}
 \item the amplitude algorithm of \sref{sec:nbaaa} and its alternative formulation in \sref{subsec:alternative_form}, and
 \item the mean estimation algorithm of \sref{sec:mean}
\end{itemize}
use an ancilla qubit to make the quantity $\Braket{\Psi_0|\Uphi|\Psi_0}$ real-valued. This is important for achieving the respective goals of the algorithms. However, the same algorithms can be performed without the ancilla, albeit to achieve different goals, which may be relevant in some use cases. In this section, the ancilla-free versions of the algorithms will be briefly described and analyzed.

\subsection{Ancilla-Free Non-Boolean Amplitude Amplification}
The ancilla-free version of the amplitude amplification algorithm is almost identical to the algorithm introduced in \sref{sec:nbaaa}. The only difference is that in the ancilla-free version, the single-register operators $S_{\psi_0}$, $U_\varphi$, and $U_\varphi^\dagger$ are used in place of the two-register operators $\Udif$, $\Uphi$, and $\Uphi^\dagger$, respectively. For concreteness, the algorithm proceeds as follows:
\begin{enumerate}
 \item Initialize a system in the state $\Ket{\psi_0}$.
 \item Act the operation $S_{\psi_0}U_\varphi$ during the odd iterations and $S_{\psi_0}U^\dagger_\varphi$ during the even iterations.
\end{enumerate}

Analogous to the two-register states $\Ket{\Alpha}$ and $\Ket{\Beta}$ in \eqref{eq:alpha} and \eqref{eq:beta}, let the single-register states $\Ket{\alpha'}$ and $\Ket{\beta'}$ be defined as
\begin{alignat}{2}
 \Ket{\alpha'} &\equiv U_\varphi \Ket{\psi_0} &&= \sum_{x=0}^{N-1} e^{i\varphi(x)}\, a_0(x)\,\Ket{x}\,,\label{eq:alpha_prime}\\
 \Ket{\beta'} &\equiv U^\dagger_\varphi \Ket{\psi_0} &&= \sum_{x=0}^{N-1} e^{-i\varphi(x)}\,a_0(x)\,\Ket{x}\,.\label{eq:beta_prime}
\end{alignat}
Analogous to $\theta$ defined in \eqref{eq:costheta}, let $\theta' \in [0, \pi/2]$ and $\delta\in [0, 2\pi)$ be implicitly defined by
\begin{equation}
 \cos(\theta')\,e^{i\delta} \equiv \Braket{\psi_0 | \alpha'} = \sum_{x=0}^{N-1} \big|a_0(x)\big|^2\,e^{i\varphi(x)}\,. \label{eq:costhetaprime}
\end{equation}
$\cos(\theta')$ and $\delta$ are the magnitude and phase, respectively, of the initial (i.e., $x$ sampled from $\Ket{\psi_0}$) expected value of $e^{i\varphi}$. An important difference between $\theta'$ and $\theta$ is that $\cos(\theta')$ is restricted to be non-negative, unlike $\cos(\theta)$, which can be positive, negative, or zero.

Note that $\cos(\theta')$ can be written as
\begin{equation}
 \cos(\theta') = \sum_{x=0}^{N-1}\big|a_0(x)\big|^2\,e^{i\varphi'(x)}\,, \label{eq:phi_prime}
\end{equation}
where $\varphi'(x)$ is given by
\begin{equation}
 \varphi'(x) \equiv \varphi(x) - \delta\,.
\end{equation}
Acting the oracle $U_\varphi$ for the function $\varphi$ can be thought of as acting the oracle $U_{\varphi'}$ for the function $\varphi'$, followed performing a global, state independent phase-shift of $\delta$. Furthermore, from \eqref{eq:phi_prime}, it can seen that $\Braket{\psi_0|U_{\varphi'}|\psi_0}$ is real-valued. This observation can be used to re-purpose the analysis in \sref{sec:anal} for the ancilla-free version; the corresponding results are presented here without explicit proofs.

Let $\Ket{\psi'_k}$ be the state of the system of the after $k\geq 0$ iterations of the ancilla-free algorithm. Analogous to \eqref{eq:Psik_full}, $\Ket{\psi'_k}$ can be written as
\begin{equation}
 \Ket{\psi'_k} = \begin{cases}
  \displaystyle\frac{e^{i\delta}}{\sin(\theta')} \Big[\sin\!\big((k+1)\theta'\big) \Ket{\psi_0} -\sin(k\theta') e^{-i\delta}\Ket{\alpha'}\Big]\,,\qquad\qquad &\text{if } k \text{ is odd}\,,\\[1.5em]
  \displaystyle\frac{1}{\sin(\theta')} \Big[\sin\!\big((k+1)\theta'\big) \Ket{\psi_0} -\sin(k\theta') e^{i\delta}\Ket{\beta'}\Big]\,,\qquad\qquad &\text{if } k \text{ is even}\,.
 \end{cases}
\end{equation}
Let $p'_K(x)$ be probability of measuring the system in state $x$ after $K$ iterations. Analogous to \eqref{eq:Pk_short}, $p'_K(x)$ can be written as
\begin{equation}
 p'_K(x) = p_0(x) \Big\{1 - \lambda'_K \Big[\cos\!\big(\varphi(x)-\delta\big) - \cos(\theta')\Big]\Big\}\,,
\end{equation}
where the $\lambda'_K$, the ancilla-free analogue of $\lambda_K$, is given by
\begin{equation}
 \lambda'_K = \frac{2\,\sin(K\theta')\,\sin\!\big((K+1)\theta'\big)}{\sin^2(\theta')} = \frac{\cos(\theta') - \cos\!\big((2K+1)\theta'\big)}{\sin^2(\theta')}\,.
\end{equation}
In this case, the probability amplification factor $p'_K/p_0$ is linear in $\cos(\varphi-\delta)$.

\subsection{Ancilla-Free Mean Estimation} \label{subsec:ancillafree_mean}
The ancilla-free mean estimation algorithm described in this subsection can estimate the \textbf{magnitude of} $\Braket{\psi_0|U|\psi_0}$ for a given unitary operator $U$. Here the algorithm is presented in terms of the oracle $U_\varphi$, and the goal of the algorithm is to estimate $\cos(\theta')$ from \eqref{eq:costhetaprime}, i.e., the magnitude of $E_{\psi_0}\!\big[e^{i\varphi}\big]\equiv \Braket{\psi_0|U_\varphi|\psi_0}$.

Let the unitary operator $\Ueo$ be defined as
\begin{equation}
 \Ueo \equiv S_{\psi_0}\,U_\varphi\,S_{\psi_0}\,U^\dagger_\varphi\,. \label{eq:ueo}
\end{equation}
Its action corresponds to performing the (ancilla-free) even-iteration operation once, followed by the odd-iteration operation. Analogous to \eqref{eq:Psi0_eta} and \eqref{eq:eta}, the state $\Ket{\psi_0}$ can be written as
\begin{equation}
 \Ket{\psi_0} = \frac{\Ket{\eta'_+} - \Ket{\eta'_-}}{\sqrt{2}}\,, \label{eq:psi0_etaprime}
\end{equation}
where $\Ket{\eta'_+}$ and $\Ket{\eta'_-}$ are given by
\begin{equation}
 \Ket{\eta'_\pm} = \frac{e^{\pm i\theta'}\Ket{\psi_0} - e^{-i\delta}\Ket{\alpha'}}{i\,\sqrt{2}\,\sin( \theta')}\,. \label{eq:etaprime_def}
\end{equation}
$\Ket{\eta'_+}$ and $\Ket{\eta'_-}$ are unit-normalized eigenstates of $\Ueo$ with eigenvalues $e^{2i\theta'}$ and $e^{-2i\theta'}$, respectively.
\begin{align}
 \Ueo\,\Ket{\eta'_\pm} &= e^{\pm 2i\theta'}\,\Ket{\eta'_\pm}\,, \label{eq:etaprime_action}\\
 \Braket{\eta'_\pm|\eta'_\pm} &= 1\,. \label{eq:etaprime_norm}
\end{align}
These properties of $\Ket{\eta'_+}$ and $\Ket{\eta'_-}$ in \eqref{eq:etaprime_action} and \eqref{eq:etaprime_norm} are proved in \aref{appendix:ueo_eigen}. The observations in \eqref{eq:psi0_etaprime} and \eqref{eq:etaprime_action} lead to the following algorithm for estimating $\cos(\theta')$:
\begin{enumerate}
 \item Perform the QPE algorithm with
 \begin{itemize}
  \item $\Ueo$ serving the role of the unitary operator under consideration, and
  \item the superposition state $\Ket{\psi_0}$ in place of the eigenstate required by the QPE algorithm as input.
 \end{itemize}
 Let the output of this step, appropriately scaled to be an estimate of the phase angle in the range $[0,2\pi)$, be $\hat{\omega}$.
 \item Return $\big|\cos\!\big(\hat{\omega}/2\big)\big|$ as the estimate for $\cos(\theta')$.
\end{enumerate}
\textbf{Proof of correctness of the algorithm:} In this version of the algorithm, $\hat{\omega}$ will be an estimate for either $2\theta$ or $2\pi-2\theta$.\footnote{If $\theta'=0$, the phase angle being estimated is $0$ for both $\Ket{\eta'_+}$ and $\Ket{\eta'_-}$.} So, $\hat{\omega}/2$ will be an estimate for either $\theta'$ or $\pi-\theta'$. Since, a) $\cos(\pi-\theta') = -\cos(\theta')$, and b) $\cos(\theta')$ is a non-negative number, it follows that $\big|\cos\!\big(\hat{\omega}/2\big)\big|$ is an estimate for $\cos(\theta')$.

\section{Summary and Outlook} \label{sec:summary}

In this paper, two new oracular quantum algorithms were introduced and analyzed. The action of the oracle $U_\varphi$ on a basis state $\Ket{x}$ is to apply a state dependent, real-valued phase shift $\varphi(x)$.

The first algorithm is the non-boolean amplitude amplification algorithm, which, starting from an initial superposition state $\Ket{\psi_0}$, preferentially amplifies the amplitudes of the basis states based on the value of $\cos(\varphi)$. In this paper, the goal of the algorithm was chosen to be to preferentially amplify the states with \emph{lower} values of $\cos\!\big(\varphi(x)\big)$. The algorithm is iterative in nature. After $K$ iterations, the probability for a measurement of the system to yield $x$, namely $p_K(x)$, differs from the original probability $p_0(x)$ by a factor that is linear in $\cos\!\big(\varphi(x)\big)$. The coefficient $-\lambda_K$ of this linear dependence controls the degree (and direction) of the preferential amplification.

The second algorithm is the quantum mean estimation algorithm, which uses QPE as a subroutine in order to estimate the expectation of $U_\varphi$ under $\Ket{\psi_0}$, i.e., $\Braket{\psi_0|U_\varphi|\psi_0}$. The algorithm offers a quadratic speedup over the classical approach of estimating the expectation, as a sample mean over randomly sampled inputs.

The non-boolean amplitude amplification algorithm and the quantum mean estimation algorithm are generalizations, respectively, of the boolean amplitude amplification and amplitude estimation algorithms. The boolean algorithms are widely applicable and feature as primitives in several quantum algorithms \cite{10.1145/261342.261346,10.1137/040605072,10.1007/BFb0054319,10.1007/s11128-017-1600-4,10.1007/978-3-540-78773-0_67,7016940,Yanhu,HOGG2000181,10.1145/301250.301349,Gong,zeng2} because of the generic nature of the tasks they accomplish. Furthermore, several extensions and variations of the boolean algorithms exist in the literature, e.g., Refs.~\cite{multiplewinning,PhysRevA.60.2742,QAAAE,toyama_dijk_nogami_2013,Yanhu,HOGG2000181, zeng1,10.5555/2600508.2600515,nisqest1,nisqest2,nisqest3,10.1137/1.9781611976014.5,nisqest4,nisqest5,yan}. 

Likewise, the non-boolean algorithms introduced in this paper also perform fairly generic tasks with a wide range of applicability. In addition, there is also a lot of scope for extending and modifying these algorithms. For example,
\begin{itemize}
 \item In this paper, the choice $\tilde{K}$ for the number of iterations to perform (in the amplitude amplification algorithm) was derived for the case where the value of $\cos(\theta)$ is a priori known. On the other hand, if $\cos(\theta)$ is not known beforehand, one can devise strategies for choosing the number of iterations as in Refs.~\cite{multiplewinning,QAAAE},
 \item There exist versions of the boolean amplification algorithm in the literature \cite{QAAAE,toyama_dijk_nogami_2013} that use a generalized version of the operator $S_{\psi_0}$ given by
 \begin{equation}
  S^\mathrm{gen}_{\psi_0}(\phi) = \big[1-e^{i\phi}\big] \Ket{\psi_0}\Bra{\psi_0} - I\,.
 \end{equation}
 The generalized operator $S^\mathrm{gen}_{\psi_0}(\phi)$ reduces to $S_{\psi_0}$ for $\phi=\pi$. Such a modification, with an appropriately chosen value of $\phi$, can be used to improve the success probability of the boolean amplification algorithm to $1$ \cite{QAAAE,toyama_dijk_nogami_2013}. A similar improvement may be possible for the non-boolean amplification algorithm as well, by similarly generalizing the $\Udif$ operator.
 \item Several variants of the (boolean) amplitude estimation algorithm exist in the literature \cite{10.5555/2600508.2600515,nisqest1,nisqest2,nisqest3,10.1137/1.9781611976014.5,nisqest4,nisqest5}, which use classical post-processing either to completely avoid using the QPE algorithm, or to reduce the depth of the circuits used in the QPE subroutine. It may be possible to construct similar variants for the mean estimation algorithm of this paper.
\end{itemize}

In the rest of this section, some assorted thoughts on the potential applications of the algorithms of this paper are presented, in no particular order.

\subsection{Approximate Optimization}
A straightforward application of the non-boolean amplitude amplification algorithm is in the optimization of objective functions defined over a discrete input space. The objective function to be optimized needs to be mapped onto the function $\varphi$ of the oracle $U_\varphi$, with the basis states of the oracle corresponding to the different discrete inputs of the objective function. After performing an appropriate number of iterations of the algorithm, measuring the state of the system will yield ``good'' states with amplified probabilities. Multiple repetitions of the algorithm (multiple shots) can be performed to probabilistically improve the quality of the optimization.

Note that the technique is not guaranteed to yield the true optimal input, and the performance of the technique will depend crucially on factors like a) the map from the objective function to the oracle function $\varphi$, b) the number of iterations $K$, c) the initial superposition $\Ket{\psi_0}$, and in particular, d) the initial distribution of $\varphi$ under the superposition $\Ket{\psi_0}$. This approach joins the list of other quantum optimization techniques \cite{HOGG2000181,PhysRevLett.121.030503}, including the Quantum Approximate Optimization Algorithm \cite{QAOA1,QAOA2} and Adiabatic Quantum Optimization \cite{FINNILA1994343,PhysRevE.58.5355,Adabatic}.

Analyzing the performance of the non-boolean amplitude amplification algorithm for the purpose of approximate optimization is beyond the scope of this work, but the results in \sref{subsec:moments} and \sref{subsec:cumulative} can be useful for such analyses.

\subsection{Simulating Probability Distributions}
The amplitude amplification algorithm could be useful for simulating certain probability distributions. By choosing the initial state $\Ket{\psi_0}$, oracle $U_\varphi$, and the number of iterations $K$, one can control the final sampling probabilities $p_K(x)$ of the basis states; the exact expression for $p_K(x)$ in terms of these factors is given in \eqref{eq:Pk_short}.

\subsection{Estimating the Overlap Between Two States} \label{subsec:overlap}
Let $\Ket{\psi}$ and $\Ket{\phi}$ be two different states produced by acting the unitary operators $A$ and $B$, respectively, on the state $\Ket{0}$.
\begin{equation}
 \Ket{\psi} = A\Ket{0},\qquad\qquad \Ket{\phi} = B\Ket{0}\,.
\end{equation}
Estimating the overlap $\big|\Braket{\psi|\phi}\big|$ between the two states is an important task with several applications \cite{PhysRevLett.87.167902,PhysRevA.69.022307,PhysRevA.95.032337,10.1145/2432622.2432625,PhysRevLett.103.150502}, including in Quantum Machine Learning (QML) \cite{Lloyd2013,PhysRevLett.113.130503,10.5555/2871393.2871400,PhysRevA.99.032331}. Several algorithms \cite{Cincio_2018,PhysRevLett.124.060503,PhysRevA.98.062318}, including the Swap test \cite{SWAP} can be used for estimating this overlap. For the Swap test, the uncertainty in the estimated value of $\big|\Braket{\psi|\phi}\big|$ falls as $\mathcal{O}(1/\sqrt{q})$ in the number of queries $q$ to the unitaries $A$ and $B$ (used to the create the states $\Ket{\psi}$ and $\Ket{\phi}$).

On the other hand, the mean estimation algorithm of this paper can also be used to estimate $\Braket{\psi|\phi}$ by noting that
\begin{equation}
 \Braket{\psi|\phi} = \Braket{0|A^\dagger\,B|0}\,.
\end{equation}
So, by setting
\begin{align}
 U&\equiv A^\dagger\,B\,,\\
 \Ket{\psi_0}&\equiv \Ket{0}\,,
\end{align}
$\Braket{\psi|\phi}$ can be estimated as $\Braket{\psi_0|U|\psi_0}$ using the mean estimation algorithm of \sref{sec:mean}. If one is only interested in the magnitude of $\Braket{\psi|\phi}$, the ancilla-free version of the mean estimation algorithm in \sref{subsec:ancillafree_mean} will also suffice. Since, for the mean estimation algorithm, the uncertainty of the estimate falls as $\mathcal{O}(1/q)$ in the number of queries $q$ to the unitaries $A$ and $B$ (or their inverses), this approach offers a \textbf{quadratic speedup} over the Swap test. Furthermore, the $\mathcal{O}(1/q)$ scaling of the error achieved by this approach matches the performance of the optimal quantum algorithm, derived in \cite{PhysRevLett.124.060503}, for the overlap-estimation task.

\subsection{Meta-Oracles to Evaluate the Superposition \texorpdfstring{$\Ket{\psi_0}$}{|psi\_0>} and Unitary \texorpdfstring{$U$}{U}} \label{subsec:metaoracle}
Recall from \eqref{eq:Psi0_eta} and \eqref{eq:eta_action} that
\begin{equation}
 \Uiter\Ket{\Psi_0} = \frac{e^{i\theta}\Ket{\Eta_+} - e^{-i\theta}\Ket{\Eta_-}}{\sqrt{2}}\,,\label{eq:psi_iter_action}
\end{equation}
where $\cos(\theta)$ is the real part of $\Braket{\psi_0|U_\varphi|\psi_0}$. Note that the parameter $\theta$ depends on the superposition $\Ket{\psi_0}$ and the unitary $U_\varphi$. The action of $\Uiter$ on $\Ket{\Psi_0}$ is to apply a phase-shift of $\theta$ on the projection along $\Ket{\Eta_+}$ and a phase-shift of $-\theta$ on the projection along $\Ket{\Eta_-}$. This property can be used to create a meta-oracle, which evaluates the superposition $\Ket{\Psi_0}$ and/or the unitary $U_\varphi$ (or a generic unitary $U$) based on the corresponding value of $\theta$. More specifically, if the circuit $A_0$ for producing $\Ket{\Psi_0}$ and/or the circuit for $U$ are additionally parameterized using ``control'' quantum registers (provided as inputs to the circuits), then a meta-oracle can be created using \eqref{eq:psi_iter_action} to evaluate the states of the control registers. The construction of such a meta-oracle is explicitly shown in \aref{appendix:metaoracle}. Such meta-oracles can be used with quantum optimization algorithms, including the non-boolean amplitude amplification algorithm of this paper, to find ``good'' values (or states) of the control registers.

Variational quantum circuits, i.e., quantum circuits parameterized by (classical) free parameters have several applications \cite{QAOA1,QAOA2,peruzzo_mcclean_shadbolt_yung_zhou_love_aspuru-guzik_o’brien_2014}, including in QML \cite{Benedetti_2019,havlíček_córcoles_temme_harrow_kandala_chow_gambetta_2019,PhysRevA.98.032309,farhiqml,10.1007/978-981-15-5619-7_8,PhysRevA.101.032308,grant_benedetti_cao_hallam_lockhart_stojevic_green_severini_2018,verdon}. Likewise, quantum circuits parameterized by quantum registers can also have applications, e.g., in QML and quantum statistical inference. The ideas of this subsection and \aref{appendix:metaoracle} can be used to ``train'' such circuits in a manifestly quantum manner; this will be explored further in future work.


\begin{acks}
The author thanks A.~Jahin, K.~Matchev, S.~Mrenna, G.~Perdue, E.~Peters for useful discussions and feedback.
%
%
The author is partially supported by the \grantsponsor{doehep}{U.S. Department of Energy, Office of Science, Office of High Energy Physics}{https://science.osti.gov/hep} QuantISED program under the grants a) ``HEP Machine Learning and Optimization Go Quantum'', Award Number \grantnum{doehep}{0000240323}, and b) ``DOE QuantiSED Consortium QCCFP-QMLQCF'', Award Number \grantnum{doehep}{DE-SC0019219}.



This manuscript has been authored by Fermi Research Alliance, LLC under Contract No. DE-AC02-07CH11359 with the U.S. Department of Energy, Office of Science, Office of High Energy Physics.
\end{acks}

\section*{Code and Data Availability}
The code and data that support the findings of this study are openly available at the following URL: \url{https://gitlab.com/prasanthcakewalk/code-and-data-availability/} under the directory named \texttt{arXiv\_2102.xxxxx}.


\printbibliography

\appendix
\section{Derivation of the System State After \texorpdfstring{$k$}{k} Iterations} \label{appendix:state_derivation}
This section contains the derivation of \eqref{eq:Psik_full} from \eqref{eq:Psik_matrix}. Let the matrix $M_\theta$ be defined as
\begin{equation}
 M_\theta \equiv \begin{bmatrix}
  2\cos(\theta) ~&~ 1\\[.5em]
  -1 ~&~ 0
 \end{bmatrix}\,.
\end{equation}
From \eqref{eq:Psik_matrix},
\begin{equation} \label{eq:Psik_matrix_M}
 \Ket{\Psi_k} = \begin{cases}
  \begin{bmatrix}
   \Ket{\Psi_0}\\[.5em]
   \Ket{\Alpha}
  \end{bmatrix}^T M_\theta^k \begin{bmatrix}
   1\\[.5em]
   0
  \end{bmatrix}\,,\qquad\qquad &\text{if } k \text{ is odd}\,,\\[2.5em]
  \begin{bmatrix}
   \Ket{\Psi_0}\\[.5em]
   \Ket{\Beta}
  \end{bmatrix}^T M_\theta^k \begin{bmatrix}
   1\\[.5em]
   0
  \end{bmatrix}\,,\qquad\qquad &\text{if } k \text{ is even}\,.
 \end{cases}
\end{equation}
where the superscript $T$ denotes transposition. $M_\theta$ can be diagonalized as
\begin{equation}
 M_\theta = S_\theta \begin{bmatrix}
  e^{-i\theta} ~&~ 0\\[.5em]
  0 ~&~ e^{i\theta}
 \end{bmatrix} S_\theta^{-1}\,, 
\end{equation}
where the matrix $S_\theta$ and its inverse $S_\theta^{-1}$ are given by
\begin{equation} \label{eq:S_Sinv}
 S_\theta = \frac{-i}{2\sin(\theta)}\begin{bmatrix}
  e^{-i\theta} ~&~ e^{i\theta}\\[.5em]
  -1 ~&~ -1
 \end{bmatrix}\,,\qquad\qquad S^{-1}_\theta = \begin{bmatrix}
  -1 ~&~ -e^{i\theta}\\[.5em]
  1 ~&~ e^{-i\theta}
 \end{bmatrix}\,.
\end{equation}
Now, $M_\theta^k$ can be written as
\begin{equation} \label{eq:Mpowk}
 M_\theta^k = S_\theta \begin{bmatrix}
  e^{-i\theta} ~&~ 0\\[.5em]
  0 ~&~ e^{i\theta}
 \end{bmatrix}^k S_\theta^{-1} = S_\theta \begin{bmatrix}
  e^{-ik\theta} ~&~ 0\\[.5em]
  0 ~&~ e^{ik\theta}
 \end{bmatrix} S_\theta^{-1}\,.
\end{equation}
From \eqref{eq:S_Sinv} and \eqref{eq:Mpowk},
\begin{subequations}
\begin{align}
 M_\theta^k \begin{bmatrix}
  1\\[.5em]
  0
 \end{bmatrix} &= S_\theta \begin{bmatrix}
  e^{-ik\theta} ~&~ 0\\[.5em]
  0 ~&~ e^{ik\theta}
 \end{bmatrix} \begin{bmatrix}
   -1\\[.5em]
   1
  \end{bmatrix} \\[.5em]
  &= S_\theta \begin{bmatrix}
   -e^{-ik\theta}\\[1em]
   e^{ik\theta}
  \end{bmatrix} = \frac{1}{\sin(\theta)} \begin{bmatrix}
   \sin\!\big((k+1)\theta\big)\\[1em]
   -\sin(k\theta)
  \end{bmatrix}\,.
\end{align}
\end{subequations}
Plugging this back into \eqref{eq:Psik_matrix_M} leads to \eqref{eq:Psik_full}.

\section{Proofs of Relevant Trigonometric Identities} \label{appendix:trig_proofs}
This section contains the derivations of the trigonometric identities \eqref{eq:trig1} and \eqref{eq:trig2}. The derivations will be based on the following standard trigonometric identities:
\begin{align}
 \sin(-C) &= -\sin(C)\,,\\
 \cos(-C) &= \cos(C)\,,\\
 \sin^2(C) + \cos^2(C) &= 1\,,\\
 \sin(C+D) &= \sin(C)\,\cos(D) + \cos(C)\,\sin(D)\,, \label{eq:sin_sum}\\
 \cos(C+D) &= \cos(C)\,\cos(D) - \sin(C)\,\sin(D)\,. \label{eq:cos_sum}
\end{align}
\subsection{Proof of \texorpdfstring{Identity \eqref{eq:trig1}}{the First Identity}}
From \eqref{eq:sin_sum},
\begin{subequations}
\begin{align}
 \begin{split}
 \sin(C+D)\,\sin(C-D) &= \Big[\sin(C)\,\cos(D) + \cos(C)\,\sin(D)\Big]\\
 &\qquad\qquad\qquad\times\Big[\sin(C)\,\cos(D) - \cos(C)\,\sin(D)\Big]
 \end{split}\\
 &= \sin^2(C)\,\cos^2(D) - \cos^2(C)\,\sin^2(D)\\
 &= \sin^2(C)\,\Big[1-\sin^2(D)\Big] - \Big[1-\sin^2(C)\Big]\,\sin^2(D)\\
 &= \sin^2(C) - \sin^2(D)\,. \label{eq:trig_troll}
\end{align}
\end{subequations}
Now, from \eqref{eq:sin_sum},
\begin{align}
 \sin^2(C) + \sin^2(C+D) &= \sin^2(C) + \sin(C+D) \Big[\sin(C)\,\cos(D) + \cos(C)\,\sin(D)\big]\\
 \begin{split}
  &= \sin^2(C) + 2\,\sin(C)\,\cos(D)\,\sin(C+D) \\
  &\qquad\qquad -\sin(C+D) \Big[\sin(C)\,\cos(D) - \cos(C)\,\sin(D)\Big]\,.\\
 \end{split}
\end{align}
Using \eqref{eq:sin_sum} again,
\begin{equation}
 \sin^2(C) + \sin^2(C+D) = \sin^2(C) + 2\,\sin(C)\,\cos(D)\,\sin(C+D) - \sin(C+D)\,\sin(C-D)\,.
\end{equation}
Using \eqref{eq:trig_troll} here leads to
\begin{equation}
 \sin^2(C) + \sin^2(C+D) = \sin^2(D) + 2\,\sin(C)\,\cos(D)\,\sin(C+D)\,,
\end{equation}
which completes the proof of \eqref{eq:trig1}.

\subsection{Proof of \texorpdfstring{Identity \eqref{eq:trig2}}{the Second Identity}}
From \eqref{eq:cos_sum},
\begin{align}
 \cos(2C+D) &= \cos(C)\,\cos(C+D) - \sin(C)\,\sin(C+D)\\
 &= \cos(-C)\,\cos(C+D) - \sin(-C)\,\sin(C+D) - 2\,\sin(C)\,\sin(C+D)
\end{align}
Using \eqref{eq:cos_sum} again leads to
\begin{align}
 \cos(2C+D) &= \cos(-C + C + D) - 2\,\sin(C)\,\sin(C+D)\\
 &= \cos(D) - 2\,\sin(C)\,\sin(C+D)\,,
\end{align}
which completes the proof of \eqref{eq:trig2}.

\section{Properties of \texorpdfstring{$\Ket{\Eta_+}$}{|eta+>} and \texorpdfstring{$\Ket{\Eta_-}$}{|eta->}} \label{appendix:uiter_eigen}
This section contains the proof of \eqref{eq:eta_action} and \eqref{eq:eta_norm}, which state that $\Ket{\Eta_+}$ and $\Ket{\Eta_-}$ are unit normalized eigenstates of $\Uiter$ with eigenvalues $e^{i\theta}$ and $e^{-i\theta}$, respectively. From \eqref{eq:uiter_psi0}, \eqref{eq:uiter_alpha}, and \eqref{eq:eta},
\begin{subequations}
\begin{align}
 \Uiter\Ket{\Eta_\pm} &= \frac{e^{\pm i\theta}\Big[2\cos(\theta)\Ket{\Psi_0} - \Ket{\Alpha}\Big] - \Ket{\Psi_0}}{i\,\sqrt{2}\,\sin(\theta)}\\
 &= e^{\pm i\theta}~\left[\frac{\Big[2\cos(\theta) - e^{\mp i\theta}\Big]\Ket{\Psi_0} - \Ket{\Alpha}}{i\,\sqrt{2}\,\sin(\theta)}\right]\\
 &= e^{\pm i\theta}~\left[\frac{e^{\pm i\theta}\Ket{\Psi_0} - \Ket{\Alpha}}{i\,\sqrt{2}\,\sin(\theta)}\right] = e^{\pm i\theta}\,\Ket{\Eta_\pm}\,.
\end{align}
\end{subequations}
This completes the proof of \eqref{eq:eta_action}. From \eqref{eq:Psi0_alpha} and \eqref{eq:eta},
\begin{subequations}
\begin{align}
 \Braket{\Eta_\pm|\Eta_\pm} &= \frac{\Braket{\Psi_0|\Psi_0} + \Braket{\Alpha|\Alpha} - e^{\pm i\theta}\Braket{\Alpha|\Psi_0} - e^{\mp i\theta}\Braket{\Psi_0|\Alpha}}{2\,\sin^2(\theta)}\\
 &= \frac{1+1-2\cos^2(\theta)}{2\sin^2(\theta)} = 1\,.
\end{align}
\end{subequations}
This completes the proof of \eqref{eq:eta_norm}.

\section{Properties of \texorpdfstring{$\Ket{\eta'_+}$}{|eta'+>} and \texorpdfstring{$\Ket{\eta'_-}$}{|eta'->}} \label{appendix:ueo_eigen}
This section contains the proof of \eqref{eq:etaprime_action} and \eqref{eq:etaprime_norm}, which state that $\Ket{\eta'_+}$ and $\Ket{\eta'_-}$ are unit normalized eigenstates of $\Ueo$ with eigenvalues $e^{2i\theta'}$ and $e^{-2i\theta'}$, respectively. From \eqref{eq:S_onereg}, \eqref{eq:alpha_prime}, \eqref{eq:beta_prime}, and \eqref{eq:costhetaprime}, the action of $\Ueo$ on $\Ket{\psi_0}$ can be written as
\begin{subequations}\label{eq:ueo_psi0}
\begin{align}
 \Ueo\Ket{\psi_0} &= S_{\psi_0}\,U_\varphi\,S_{\psi_0}\Ket{\beta'}\\
 &= S_{\psi_0}\,U_\varphi\Big[2\Braket{\psi_0|\beta'}\Ket{\psi_0} - \Ket{\beta'}\Big]\\
 &= S_{\psi_0}\Big[2\Braket{\psi_0|\beta'}\Ket{\alpha'} - \Ket{\psi_0}\Big]\\
 &= 2\Braket{\psi_0|\beta'}\Big[2\Braket{\psi_0|\alpha'}\Ket{\psi_0} - \Ket{\alpha'}\Big] - \Ket{\psi_0}\\
 &= \big[4\cos^2(\theta')-1\big]\Ket{\psi_0} - 2\cos(\theta')e^{-i\delta}\Ket{\alpha'}\,,
\end{align}
\end{subequations}
and the action of $\Ueo$ on $\Ket{\alpha'}$ can be written as
\begin{subequations}\label{eq:ueo_alphaprime}
\begin{align}
 \Ueo\Ket{\alpha'} &= S_{\psi_0}\,U_\varphi\,S_{\psi_0}\Ket{\psi_0}\\
 &= S_{\psi_0}\,U_\varphi\Ket{\psi_0}\\
 &= S_{\psi_0}\Ket{\alpha'}\\
 &= 2\Braket{\psi_0|\alpha'}\Ket{\psi_0} - \Ket{\alpha'}\\
 &= 2\cos(\theta')e^{i\delta}\Ket{\psi_0} - \Ket{\alpha'}\,.
\end{align}
\end{subequations}
Now, using \eqref{eq:ueo}, \eqref{eq:ueo_psi0}, and \eqref{eq:ueo_alphaprime},
\begin{equation}
 \Ueo\Ket{\eta'_\pm} = \frac{\Big[4\cos^2(\theta')-1 - 2\cos(\theta')e^{\mp i\theta'}\Big]e^{\pm i\theta'}\Ket{\psi_0} - \Big[2\cos(\theta')e^{\pm i\theta'}-1\Big]e^{-i\delta}\Ket{\alpha'}}{i\,\sqrt{2}\,\sin( \theta')}\,.\label{eq:ueo_proof1}
\end{equation}
Using $2\cos(\theta') = e^{i\theta'} + e^{-i\theta'}$ it can be shown that
\begin{equation}
 4\cos^2(\theta')-1 - 2\cos(\theta')e^{\mp i\theta'} = 2\cos(\theta')e^{\pm i\theta'}-1 = e^{\pm 2i\theta'}\,.
\end{equation}
Using this identity, \eqref{eq:ueo_proof1} can be simplified as
\begin{equation}
 \Ueo\Ket{\eta'_\pm} = e^{\pm 2i\theta'}\Ket{\eta'_\pm}\,.
\end{equation}
This completes the proof of \eqref{eq:etaprime_action}.

From \eqref{eq:costhetaprime} and \eqref{eq:etaprime_def},
\begin{subequations}
\begin{align}
 \Braket{\eta'_\pm|\eta'_\pm} &= \frac{\Braket{\psi_0|\psi_0} + \Braket{\alpha'|\alpha'}-e^{\pm i\theta'+i\delta}\Braket{\alpha'|\psi_0} - e^{\mp i\theta'-i\delta}\Braket{\psi_0|\alpha'}}{2\sin^2(\theta')}\\
 &= \frac{1+1- e^{\pm i\theta'}\cos(\theta') - e^{\mp i\theta'}\cos(\theta')}{2\sin^2(\theta')} = \frac{2 - 2\cos^2(\theta')}{2\sin^2(\theta')} = 1\,.
\end{align}
\end{subequations}
This completes the proof of \eqref{eq:etaprime_norm}.

\section{Meta-Oracle Construction} \label{appendix:metaoracle}
This section describes the construction of the meta-oracle discussed in \sref{subsec:metaoracle}. For the purposes of this section, it is more natural to work with generic unitary operators $U$, instead of oracles $U_\varphi$ with a known eigenbasis. Accordingly, the subscript ``$\varphi$'' will be dropped from the oracles $U_\varphi$ and $\Uphi$. Similarly, the subscript ``$0$'' will be dropped from the circuit $A_0$, and the states $\Ket{\psi_0}$ and $\Ket{\Psi_0}$, since they are not to be interpreted as ``initial states'' in this section. The symbol $\Ket{\cdot}$ will refer to an arbitrary pure state of a quantum system.

\begin{figure}[t]
 \centering
 \begin{tikzpicture}
  \node[scale=1.0] {
   \begin{quantikz}
    \lstick{$MA$ control\\$\Ket{x_A}_\mathrm{ctrl}$} & \ctrlbundle{1} & \qwmult \rstick{~} \\[1em]
    \lstick{} & \gate{A(x_A)}\qwmult & \qwmult \rstick{~}
   \end{quantikz}
  };
 \end{tikzpicture}\qquad\qquad
 \begin{tikzpicture}
  \node[scale=1.0] {
   \begin{quantikz}
    \lstick{$MU$ control\\$\Ket{x_U}_\mathrm{ctrl}$} & \ctrlbundle{1} & \qwmult \rstick{~} \\[1em]
    \lstick{} & \gate{U(x_U)}\qwmult & \qwmult \rstick{~}
   \end{quantikz}
  };
 \end{tikzpicture}
 \caption{Meta-circuits $MA$ (left panel) and $MU$ (right panel), which implement the parameterized operations $A(x_A)$ and $U(x_U)$, respectively. The actions of the circuits $MA$ and $MU$ are shown for the case when their control registers are in the basis states $\Ket{x_A}_\mathrm{ctrl}$ and $\Ket{x_U}_\mathrm{ctrl}$, respectively.}
 \label{fig:metaAU}
\end{figure}
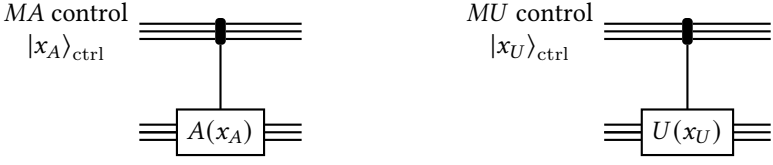

Let $MA$ be a meta-circuit for preparing the state $\Ket{\psi}$, parameterized by an additional ``control'' quantum register (with an orthonormal basis $\{\Ket{0}_\mathrm{ctrl},\Ket{1}_\mathrm{ctrl},\dots,\Ket{N_A-1}_\mathrm{ctrl}\}$) as follows:
\begin{alignat}{3}
 MA \Big[\Ket{x_A}_\mathrm{ctrl}&\otimes \Ket{\cdot}\Big] &&= \Ket{x_A}_\mathrm{ctrl}\otimes \big[A(x_A)\Ket{\cdot}\big]\,,\qquad &&\forall x_A\in\{0,\dots,N_A-1\}\,,\\
 MA \Big[\Ket{x_A}_\mathrm{ctrl}&\otimes \Ket{0}\Big] &&= \Ket{x_A}_\mathrm{ctrl}\otimes \Ket{\psi(x_A)}\,,\qquad &&\forall x_A\in\{0,\dots,N_A-1\}\,.
\end{alignat}
Here, the operator $A(x_A)$ and the state $\Ket{\psi(x_A)}$ are both parameterized by $x_A$ via the quantum register with the subscript ``$\mathrm{ctrl}$''. Likewise, let $MU$ be a meta-circuit for the unitary $U$, parameterized by an additional quantum register (with an orthonormal basis $\{\Ket{0}_\mathrm{ctrl},\dots,\Ket{N_U-1}_\mathrm{ctrl}\}$) as follows:
\begin{align}
 MU \Big[\Ket{x_U}_\mathrm{ctrl}\otimes \Ket{\cdot}\Big] = \Ket{x_U}_\mathrm{ctrl}\otimes \big[U(x_U)\Ket{\cdot}\big]\,,\qquad \forall x_U\in\{0,\dots,N_U-1\}\,.
\end{align}
Here $U(x_U)$ is parameterized by the parameter $x_U$. The unitary operations performed by $MA$ and $MU$ can be written as
\begin{align}
 MA = \sum_{x_A=0}^{N_A-1} \Big[\Ket{x_A}\Bra{x_A}\Big]_\mathrm{ctrl} \otimes A(x_A)\,,\\
 MU = \sum_{x_U=0}^{N_U-1} \Big[\Ket{x_U}\Bra{x_U}\Big]_\mathrm{ctrl} \otimes U(x_U)\,.
\end{align}
The meta-circuits $MA$ and $MU$ are depicted in \fref{fig:metaAU}.
Under this setup,
\begin{itemize}
 \item The two-register state $\Ket{\Psi}$ and operator $\S_{\Psi}$ will be parameterized as
 \begin{align}
  \Ket{\Psi(x_A)} &= \Ket{+}\otimes \Ket{\psi(x_A)}\,,\\
  \S_{\Psi}(x_A) &= 2\Ket{\Psi(x_A)}\Bra{\Psi(x_A)} - \I\,.
 \end{align}
 \item The two-register operator $\U = H\otimes U$ will be parameterized as $\U(x_U)$.
 \item The quantity $\theta$ and the two-register states $\Ket{\Eta_\pm}$ will be parameterized as
 \begin{align}
  \theta(x_A, x_U) &= \arccos{\Big[\RE\big[\Braket{\psi(x_A)|U(x_U)|\psi(x_A}\big]\Big]}\,,\\
  \Ket{\Eta_\pm(x_A, x_U)} &= \frac{e^{\pm i\theta(x_A, x_U)}\Ket{\Psi(x_A)} - \U(x_U)\Ket{\Psi(x_A)}}{i\,\sqrt{2}\,\sin\!\big(\theta(x_A, x_U)\big)}\,.
 \end{align}
\end{itemize}

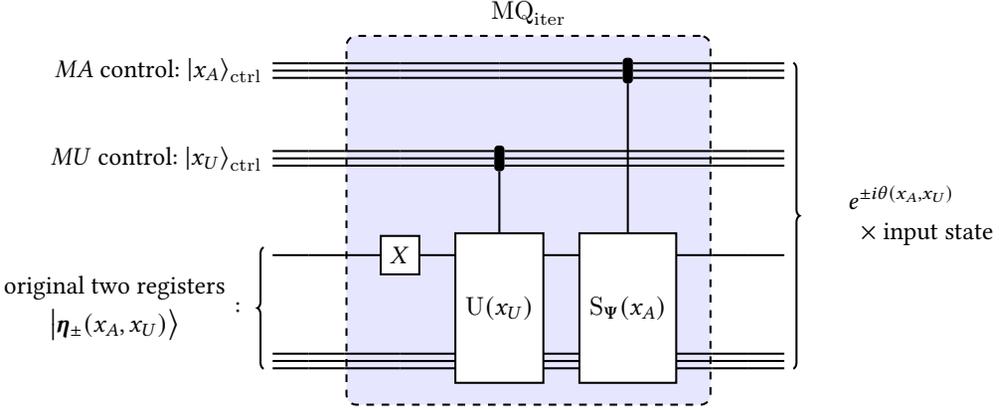
\begin{figure}[t]
 \centering
 \resizebox{\textwidth}{!}{
  \begin{tikzpicture}
   \node[scale=1.0] {
    \begin{quantikz}
     \lstick{$MA$ control: $\Ket{x_A}_\mathrm{ctrl}$} & \qwmult & \qwmult & \qwmult \gategroup[wires=4,steps=3,style={dashed,rounded corners,fill=blue!10,inner xsep=10pt,inner ysep=5pt,outer ysep=3pt},background]{$\mathbf{MQ}_\mathrm{iter}$} & \qwmult & \ctrlbundle{2} & \qwmult & \qwmult & \qwmult \rstick[wires=4]{$e^{\pm i\theta(x_A, x_U)}$\\\qquad$\times$ input state} \\[1em]
     \lstick{$MU$ control: $\Ket{x_U}_\mathrm{ctrl}$} & \qwmult & \qwmult & \qwmult & \ctrlbundle{2} & \qwmult & \qwmult & \qwmult & \qwmult \\[1em]
     \lstick[wires=2]{$\displaystyle\genfrac{}{}{0pt}{}{\text{original two registers}}{\Ket{\big.\Eta_\pm(x_A, x_U)}}$ : } & \qw & \qw & \gate{X} & \gate[wires=2]{\U(x_U)} & \gate[wires=2]{\S_{\Psi}(x_A)} & \qw & \qw & \qw \\[1em]
     & \qwmult & \qwmult & \qwmult & \qwmult & \qwmult & \qwmult & \qwmult & \qwmult
   \end{quantikz}
   };
  \end{tikzpicture}
 }
 \caption{Quantum circuit for the meta-oracle $\mathbf{MQ}_\mathrm{iter}$, which evaluates the states of the control registers. The action of the oracle is shown for the case when the inputs state is follows: The control registers of $MA$ and $MU$ are in basis states $\Ket{x_A}_\mathrm{ctrl}$ and $\Ket{x_U}_\mathrm{ctrl}$, respectively, and the original two (non-control) registers are in the state $\Ket{\Eta_\pm(x_A, x_U)}$.}
 \label{fig:metaoracle}
\end{figure}

Now, using $MA$ and $MU$, one can create a meta-operator $\mathbf{MQ}_\mathrm{iter}$, which is simply $\Uiter$ parameterized additionally by $x_A$ and $x_U$. The circuit for $\mathbf{MQ}_\mathrm{iter}$ is shown in \fref{fig:metaoracle}.
From \eqref{eq:eta_action}, the action of $\mathbf{MQ}_\mathrm{iter}$ on $\Ket{\Eta_\pm(x_A, x_U)}$ can be written as
\begin{equation}
 \mathbf{MQ}_\mathrm{iter} \Big[\Ket{x_A,x_U}_\mathrm{ctrl} \otimes\Ket{\Eta_\pm(x_A, x_U)}\Big] = e^{\pm i\theta(x_A, x_U)} \Big[\Ket{x_A,x_U}_\mathrm{ctrl} \otimes\Ket{\Eta_\pm(x_A, x_U)}\Big]\,.
\end{equation}
From this equation, it can be see that $\mathbf{MQ}_\mathrm{iter}$ acts as a meta-oracle that evaluates the state of the controls registers based on the corresponding value of $\theta$. Furthermore, $\mathbf{MQ}_\mathrm{iter}$ accomplishes this task using only $\mathcal{O}(1)$ calls to $MA$ and $MU$. This meta-oracle can be used with the non-boolean amplitude amplification algorithm of this paper to find ``good'' states for the control registers. Note that to use $\mathbf{MQ}_\mathrm{iter}$ as an oracle for the control registers, the original two (non-control) registers must be coupled to the control registers using $A(x_A)$:
\begin{equation}
 \Ket{+}\otimes \Big[A(x_A)\Ket{0}\Big] = \Ket{\Psi(x_A)} = \frac{\Ket{\Eta_+(x_A, x_U)} - \Ket{\Eta_-(x_A, x_U)}}{\sqrt{2}}\,.
\end{equation}
This coupling step can be incorporated into the circuit used to create the initial superposition input of the meta-oracle $\mathbf{MQ}_\mathrm{iter}$.

A construction similar to the one in this section can be used to create a meta-oracle $M\Ueo$ (parameterized version of $\Ueo$), which evaluates the control registers based on
\begin{equation}
 \cos\!\big(\theta'(x_A, x_U)\big) = \big|\Braket{\psi(x_A)|U(x_U)|\psi(x_A)}\big|\,.
\end{equation}

\end{document}